\documentclass[aps,prd,reprint,nofootinbib,longbibliography]{revtex4-1}

\usepackage{amsmath}
\usepackage{mathtools}
\usepackage{graphicx}
\usepackage[normalem]{ulem}
\usepackage[com pat=1.1.0]{tikz-feynman}
\usepackage{tikz}
\usepackage{aas_macros}
\usetikzlibrary{external}
\usepackage{comment}
\usetikzlibrary{arrows}
\usepackage[colorlinks=true, allcolors=blue]{hyperref}

\newcommand\barparena[1]{\overset{%
   \scriptscriptstyle(-)}{#1}}

\begin{document}

\title{Connecting small-scale to large-scale structures of fast neutrino-flavor conversion}

\author{Hiroki Nagakura}
\email{hiroki.nagakura@nao.ac.jp}
\affiliation{Division of Science, National Astronomical Observatory of Japan, 2-21-1 Osawa, Mitaka, Tokyo 181-8588, Japan}
\author{Masamichi Zaizen}
\affiliation{Faculty of Science and Engineering, Waseda University, Tokyo 169-8555, Japan}

\begin{abstract}
We present a systematic study of fast neutrino-flavor conversion (FFC) with both small-scale and large-scale numerical simulations in spherical symmetry. We find that FFCs can, in general, reach a quasi-steady state, and these features in the non-linear phase are not characterized by the growth rate of FFC instability but rather angular structures of electron neutrino lepton number (ELN) and heavy one (XLN). Our result suggests that neutrinos can almost reach a flavor equipartition even in cases with low growth rate of instability (e.g., shallow ELN crossing) and narrow angular regions (in momentum space) where flavor conversions occur vigorously. This exhibits that ELN and XLN angular distributions can not provide a sufficient information to determine total amount of flavor conversion in neutrinos and antineutrinos of all flavors. Based on the results of our numerical simulations, we provide a new approximate scheme of FFC that is designed so that one can easily incorporate effects of FFCs in existing classical neutrino transport codes for the study of core-collapse supernova (CCSN) and binary neutron star merger (BNSM). The scheme has an ability to capture key features of quasi-steady state of FFCs without solving quantum kinetic neutrino transport, which will serve to facilitate access to FFCs for CCSN and BNSM theorists.
\end{abstract}
\maketitle

\section{Introduction}\label{sec:intro}
There is mounting evidence that neutrinos undergo flavor conversion. Occurrences of flavor conversion imply that neutrinos have multiple eigenstates with different masses, and their mass eigenstates do not coincide with the flavor ones. In the framework of three flavors, the mass and flavor state can be connected through the Pontecorvo-Maki-Nakagawa-Sakata (PMNS) matrix with three different mixing angles and a CP-violation phase. Determining the mixing parameters and the mass differences of neutrinos is a fundamental problem in particle physics, which is also of great importance in understanding astrophysical phenomena involving neutrinos.

Neutrino flavor conversions can be interpreted through the dispersion relation (DR). The disparity of DR due to different masses of neutrinos leads to the oscillation of flavors while propagating in vacuum. When neutrinos propagate in a medium, the DR is modified by refractive effects due to coherent forward scatterings by matter. This triggers large flavor conversions if refractive effects resonate with vacuum oscillation, which is known as Mikheyev-Smirnov-Wolfenstein (MSW) resonance \cite{1979PhRvD..20.2634W,1989PrPNP..23...41M}. On the other hand, coherent forward scatterings of neutrinos themselves also provide another refractive effects \cite{1992PhLB..287..128P}. One noticeable feature is that off-diagonal components of self-interaction Hamiltonian are in general not zero on the flavor basis, which is qualitatively different from matter potential, and more importantly the flavor conversion occurs non-linearly. This generates rich phenomenologies of neutrino dynamics including collective neutrino oscillations (see, e.g., \citep{2010ARNPS..60..569D}).

Fast neutrino-flavor conversion (FFC) has attracted a great deal of attention recently \cite{2016NuPhB.908..366C,2020arXiv201101948T,2022Univ....8...94C,2022arXiv220703561R}. FFC does not depend on differences of neutrino mass, but rather being solely dictated by neutrino self-interactions. Recent studies of FFCs by linear stability analysis \cite{2017PhRvL.118b1101I}, its surrogate methods \cite{2018PhRvD..98j3001D,2020JCAP...05..027A,2021PhRvD.103l3012J,2021PhRvD.104f3014N,2022arXiv220608444R}, and some toy models (e.g., \cite{2017PhRvD..95j3007W}) have suggested that FFC instabilities would occur in CCSN \cite{2019ApJ...886..139N,2019ApJ...883...80S,2020PhRvD.101b3018D,2020PhRvR...2a2046M,2021PhRvD.104h3025N,2021PhRvD.103f3033A,2021PhRvD.103f3013C,2022ApJ...924..109H} and in BNSM \cite{2017PhRvD..95j3007W,2017PhRvD..96l3015W,2020PhRvD.102j3015G,2021JCAP...01..017P,2021PhRvL.126y1101L,2022PhRvD.105h3024J,2022arXiv220608444R}. The growth timescale of the instability can be an order of sub-nanoseconds, which is much shorter than any scale on interest in these systems. This exhibits that FFCs may radically change the neutrino radiation fields in CCSN and BNSM.

Increasing the possibility of occurrences of FFCs provided the impetus to link FFCs to theoretical models of CCSN and BNSM. The ab-initio approach to incorporate effects of FFC into CCSN/BNSM simulations requires solving seven-dimensional (1 in time, 3 in space, and 3 in momentum space) quantum kinetic equation (QKE). This poses a formidable computational challenge, however. The wavelength of neutrino oscillation is several orders of magnitude smaller than the scale height of the fluid flow, making these simulations intractable under currently available computational resources. Thus far, previous studies have mostly concentrated their efforts on local simulations \cite{2017JCAP...02..019D,2019PhRvD.100b3016M,2020PhRvD.102j3017J,2020PhLB..80035088M,2020PhRvD.102f3018B,2021PhRvL.126f1302B,2021PhRvD.104j3003W,2021PhRvD.103f3001M,2021PhRvD.104j3023R,2021PhRvD.104h3035Z,2021PhRvD.104l3026D,2022JCAP...03..051A,2022PhRvD.105d3005S,2022arXiv220505129B,2022PhRvD.106d3011R,2022arXiv220702214G} (but see also other efforts for global simulations \cite{2021JCAP...01..017P,2022arXiv220600676S,2022arXiv220704058S}), or on neutrino-radiation hydrodynamic simulations with phenomenological models of FFCs \cite{2021PhRvL.126y1101L,2022PhRvD.105h3024J,2022arXiv220710680F}. In the latter approach, the neutrino transport is essentially a classical treatment, but effects of flavor conversion are incorporated by some neutrino-mixing prescriptions. More specifically, they impose a certain condition to detect occurrences of FFCs (which is based on stability analyses), and then the total amount of flavor conversion is determined with a parametric way (or assumed to be a flavor equipartition). These simulations are useful to demonstrate how FFCs give impacts on CCSN and BNSM dynamics qualitatively. On the other hand, their outcome hinges on the instability criteria and the choice of parameter for neutrino mixings, exhibiting that better approximate prescriptions are required to gauze accurate sensitivity of CCSN and BNSM dynamics to FFCs.

Recently we proposed a novel approach to pave the way towards incorporating FFCs into CCSN and BNSM simulations \cite{2022arXiv220604097N} (hereafter the paper is referred to as NZv1). In this approach, neutrino transport is solved with quantum kinetic treatments with attenuating neutrino Hamiltonian potentials parametrically. Thanks to the attenuation of the Hamiltonian, large-scale FFC simulations can be carried out with feasible computational costs. It is also worth to note that our proposed method can be used for other studies of neutrino flavor conversions; for instance, \citet{2022arXiv221008254X} recently carried out large-scale simulations of collisional instability with attenuating Hamiltonian.

In NZv1, we performed FFC simulations in $50 {\rm km}$ spatial scales ($50 {\rm km} \le R \le 100 {\rm km}$), and then we analyzed their global features. We found that the time-averaged neutrino distributions are insensitive to the attenuation of Hamiltonian\footnote{A word of caution should be spent here. Extreme attenuation of Hamiltonian potential lead to no flavor conversion. This indicates that there is a threshold in the attenuation-parameter to capture the qualitative trend of FFCs in global scales.}, suggesting that the similar time-averaged profile would appear in the case without the attenuation. We also found in NZv1 that the difference of angular distributions of ELN (electron-neutrino lepton number) and XLN (heavy-neutrino lepton number) is a key quantity to determine the non-linear saturation of flavor conversion, and to characterize the subsequent quasi-steady state of FFCs. In fact, the ELN-XLN angular crossings become very shallow or even disappear in the time-averaged profile after the system reaches non-linear saturation. As such, NZv1 illustrated that the proposed method, attenuating Hamiltonian, can bring new insights on FFCs. This method is also expected to play a crucial role to connect local- and global features of neutrino quantum kinetics.

In this paper, we extend our previous study in NZv1 by covering various initial states of neutrinos. This study is motivated by the fact that we focused on the ability of our new approach in NZv1, and therefore we fixed the initial angular distributions of neutrinos. However, it is necessary to carry out a systematic study for various initial conditions so as to capture generic features of FFCs. To analyze the large-scale numerical simulations, we also carry out local simulations in the vicinity of inner boundary without attenuation of Hamiltonian. We shall show that some intrinsic features of FFCs can be complemented from these small-scale simulations. Finally, we provide an approximate method that determines quasi-steady states of FFC without solving QKE. For future users, we provide a recipe of the method, which can be easily implemented in existing classical neutrino transport codes.

This paper is structured as follows. In Sec.~\ref{sec:method} we first review the essence of our approach, attenuation of Hamiltonian potentials, for large-scale QKE simulations. We then describe our models in Sec.~\ref{sec:model}. All numerical results presented in this paper are encapsulated in Sec.~\ref{sec:result}. The approximate method to determine the quasi-steady state of FFCs is described in Sec.~\ref{sec:phenomodel}. Finally, we summarize our conclusions and key messages from the present work in Sec.~\ref{sec:summary}. Throughout the paper, we use the unit with $c = \hbar = 1$, where $c$ and $\hbar$ are the light speed and the reduced Planck constant, respectively; we choose the metric signature of $- + + +$.

\begin{table*}[t]
\caption{Set of parameters in our models. See Eqs.~\ref{eq:BasEq}-~\ref{eq:defalpha} for the definition of $\alpha$, $\barparena{\beta}_{ee}$, and $\xi$. $\Delta R$ corresponds to the width of computational domain. We note that the inner boundary is located at $R= 50 {\rm km}$ for all models (including local simulations). $N_r$ and $N_{\theta_{\nu}}$ represent the number of grid points in space and in neutrino angular directions, respectively. We employ uniform grids for each direction. ${\rm T}_{\rm sim}$ corresponds to the physical time of simulation. The values in parentheses in the same column (but only for reference models) denote ${\rm T}_{\rm sim}$ for extended simulations to analyze temporal variations of FFC by Fourier analysis; see text for more details.
}
\begin{tabular}{cccccccccc} \hline
~~model~~ & ~~$\alpha$~~ & ~~$\beta_{ee}$~~ & ~~$\bar{\beta}_{ee}$~~ & ~~$\xi$~~ & ~~$\Delta R$ [km] ~~  & ~~$N_r$~~ & ~~$N_{\theta_{\nu}}$~~ & ~~${\rm T}_{\rm sim}$ [ms] ~~ \\
 \hline \hline
GL-Ref & 1  & 0 & 1 & $2 \times 10^{-4}$ & 50 & 49152 & 128 & $0.5 (1)$ \\
LO-Ref & 1  & 0 & 1 & $1$ & $10^{-2}$ & 49152 & 128 & $10^{-4} (2 \times 10^{-4})$ \\
GL-$\alpha$09 & 0.9  & 0 & 1 & $2 \times 10^{-4}$ & 50 & 49152 & 128 & $0.5$ \\
LO-$\alpha$09 & 0.9  & 0 & 1 & $1$ & $10^{-2}$ & 49152 & 128 & $10^{-4}$ \\
GL-$\alpha$11 & 1.1  & 0 & 1 & $2 \times 10^{-4}$ & 50 & 49152 & 128 & $0.5$ \\
LO-$\alpha$11 & 1.1  & 0 & 1 & $1$ & $10^{-2}$ & 49152 & 128 & $10^{-4}$ \\
GL-$\bar{\beta}$01$\xi$-3 & 1 & 0 & 0.1 & $2 \times 10^{-3}$ & 50 & 49152 & 128 & $0.5$ \\
LO-$\bar{\beta}$01 & 1 & 0 & 0.1 & $1$ & $10^{-2}$ & 49152 & 128 & $10^{-4}$ \\
GL-$\bar{\beta}$001$\xi$-2 & 1 & 0 & $10^{-2}$ & $2 \times 10^{-2}$ & 50 & 49152 & 128 & $0.5$ \\
LO-$\bar{\beta}$001 & 1 & 0 & $10^{-2}$ & $1$ & 0.1 & 49152 & 128 & $10^{-3}$ \\
GL-$\bar{\beta}$0001$\xi$-1 & 1 & 0 & $10^{-3}$ & $2 \times 10^{-1}$ & 50 & 49152 & 128 & $0.5$ \\
GL-H-$\bar{\beta}$0001$\xi$-1 & 1 & 0 & $10^{-3}$ & $2 \times 10^{-1}$ & 50 & 98304 & 256 & $0.5$ \\
LO-$\bar{\beta}$0001 & 1 & 0 & $10^{-3}$ & $1$ & 1 & 49152 & 128 & $10^{-2}$ \\
GL-$\beta$05$\xi$-3 & 1  & 0.5 & 1 & $2 \times 10^{-3}$ & 50 & 49152 & 128 & $0.5$ \\
LO-$\beta$05 & 1  & 0.5 & 1 & $1$ & $10^{-2}$ & 49152 & 128 & $10^{-4}$ \\
GL-Flip & 1  & 1 & 0 & $2 \times 10^{-4}$ & 50 & 49152 & 128 & $0.5$ \\
LO-Flip & 1  & 1 & 0 & $1$ & $10^{-2}$ & 49152 & 128 & $10^{-4}$ \\
 \hline
\end{tabular}
\label{tab:model}
\end{table*}

\section{Method}\label{sec:method}
The numerical simulations presented in this paper are carried out with a newly developed QKE neutrino transport code, GRQKNT. Details of the design and a suite of tests are presented in \cite{2022PhRvD.106f3011N}. Here, we describe only the essential components of the code directly related to this present work.

In GRQKNT, we adopt a discrete-ordinate Sn method. The transport operator is handled with 5th-order weighted essentially non-oscillatory (WENO) scheme with a five-stage fourth-order TVD Runge-Kutta. In this study, we assume spherical symmetry and ignore general relativistic effects, fluid-velocity dependence, and the collision term. The resultant QKE can be written as,
\begin{equation}
  \begin{split}
& \frac{\partial \barparena{f}}{\partial t}
+ \frac{1}{r^2} \frac{\partial}{\partial r} ( r^2 \cos \theta_{\nu}  \barparena{f} )  - \frac{1}{r \sin \theta_{\nu}\
} \frac{\partial}{\partial \theta_{\nu}} ( \sin^2 \theta_{\nu} \barparena{f}) \\
&  = - i \hspace{0.5mm} \xi \hspace{0.5mm} [\barparena{H},\barparena{f}],
  \end{split}
\label{eq:BasEq}
\end{equation}
where $f$ and $\bar{f}$ represent the density matrix of neutrinos and antineutrinos, respectively. $t, r,$ and $\theta_{\nu}$ denote time, radius, and neutrino flight angle with respect to radial direction, respectively. $H$ ($\bar{H}$) represents the neutrino (antineutrino) oscillation Hamiltonian potential, which is composed of vacuum-, matter-, and self-interaction components. In this study, the matter potential is set to be zero, but we reduce the mixing angle in the vacuum potential from that constrained by experiments. This is a common prescription to effectively include effects of matter potential\footnote{It is also equivalent to work with polarization vectors of neutrinos in a co-rotating frame, see \cite{2006PhRvD..74l3004D}.}. In this study, the vacuum potential is added as a perturbation to trigger FFCs\footnote{As we shall show in Sec.~\ref{sec:result}, flavor conversions are affected by vacuum potentials in some of our models. Although the neutrino dynamics in these models are not purely dictated by the instability of FFC, their results are interesting because some interactions between fast and slow modes emerge. See Sec.~\ref{sec:result} for more details.}. Following the previous studies as NZv1, we adopt the two-flavor approximation with $\Delta m^2 = 2.5 \times 10^{-6} {\rm eV^2}$, $\theta_{\rm mix} = 10^{-6}$ and $E_{\nu}=12 {\rm MeV}$, where $\Delta m^2$ and $\theta_{\rm mix}$ denote a squared mass difference of neutrinos, mixing angle, and neutrino energy, respectively. We solve QKE on a single neutrino-energy bin, i.e., adopting monochromatic energy approximation. This is a reasonable treatment for FFCs, unless energy-dependence of neutrino-matter interactions (i.e., collision term) has an influence on flavor conversion \cite{2022PhRvD.106l3013K}.

$\xi$ ($0 \le \xi \le 1$) in the right hand side of Eq.~\ref{eq:BasEq} is not a physical quantity. It is a parameter that controls the attenuation of all neutrino oscillation Hamiltonian (vacuum, matter, and self-interaction components). When we set $\xi=1$ (for local simulations), this restores the original QKE equation. On the other hand, we set $\xi$ to be less than unity so as to make large-scale simulations ($> 10 {\rm km}$) tractable. In these simulations, spurious evolutions of FFC inevitably arise, but these unphysical features sensitively depend on $\xi$, indicating that we can identify these artifacts by convergence study with respect to $\xi$. In NZv1, we performed such a convergence study and demonstrated how physically meaningful features can be extracted from these simulations. Another thing we do notice here is that results of local simulations help us to understand those of large-scale simulations, which will be demonstrated in Sec.~\ref{sec:result}.

\section{Model}\label{sec:model}

Numerical setup in the present study is designed so as to emulate situations in the core of CCSN and BNSM. In this study, we pay special attention to FFCs driven by neutrinos propagating outwards ($\cos \theta_{\nu } \ge 0$) outside of neutrino sphere ($50 {\rm km} \le R \le 100 {\rm km}$). According to recent theoretical studies (see, e.g., \cite{2021PhRvL.126y1101L,2022PhRvD.105h3024J,2019ApJ...886..139N,2021PhRvD.104h3025N,2022arXiv220608444R,2022ApJ...924..109H}), ELN crossings likely appear in these regions, that exhibits the sign of occurrences of FFCs. It should be mentioned that FFCs can also occur inside of neutrino spheres; for instances, optically thick region \cite{2019ApJ...886..139N,2020PhRvD.101b3018D,2020PhRvD.101f3001G,2021PhRvD.104h3025N} and in semi-transparent one \cite{2022arXiv220600676S,2022arXiv220704058S}. In these regions, neutrino angular distributions are nearly isotropic, indicating that neutrinos propagating in all angles have non-negligible contributions on the self-interaction Hamiltonian potential. In addition to this, the interplay between FFCs and neutrino-matter interactions would lead to more complex dynamics in neutrino radiation field. Addressing the issue of FFCs inside the neutrino sphere is a beyond the scope of this paper, and the detailed investigation will be made in a separate paper.

In our models, we set $R_{in}=50 {\rm km}$, where $R_{in}$ denotes the radius of inner boundary. We adopt a Dirichlet boundary condition for outgoing neutrinos, and their angular distributions of $\nu_e$ and $\bar{\nu}_e$ are determined with the following equation,
\begin{equation}
\barparena{f}_{ee} = \langle \barparena{f}_{ee}\rangle \biggl( 1 + \barparena{\beta}_{ee} ( \cos \theta_{\nu} - 0.5 ) \biggr) \hspace{4mm} \cos \theta_{\nu} \ge 0.
\label{eq:iniang_in}
\end{equation}
For the sake of simplicity, other components of density matrix are set to be zero. For incoming neutrinos ($\cos \theta_{\nu} < 0$), we use a free-streaming boundary condition, which is appropriate to meet the causality requirement. At the outer boundary, on the other hand, we adopt the Dirichlet boundary condition for incoming neutrinos, which are
\begin{equation}
\barparena{f}_{ee} = \langle \barparena{f}_{ee} \rangle \times \eta \hspace{4mm} \cos \theta_{\nu} < 0.
\label{eq:iniang_out}
\end{equation}
$\eta$ represents the diluteness of incoming neutrinos, and we set $\eta=10^{-6}$. Due to the small number of incoming neutrinos, they do not contribute the self-interaction potential. Similar as the inner boundary condition, we adopt a free-streaming boundary condition for outgoing neutrinos to be consistent with the causality.

We construct angular distributions of $\nu_e$ and $\bar{\nu}_e$ by setting four parameters: two for $\langle \barparena{f}_{ee}\rangle$ and two for $\barparena{\beta}_{ee}$ in Eqs.~\ref{eq:iniang_in}~and~\ref{eq:iniang_out}. $\langle \barparena{f}_{ee}\rangle$ is a parameter to characterize the number density of $\nu_e$ ($n_{\nu_e}$) and $\bar{\nu}_e$ ($n_{\bar{\nu}_e}$) at the inner boundary. In this study, we fix $\langle f_{ee}\rangle$ in all models so that $n_{\nu_e}$ becomes $6 \times 10^{32} {\rm cm^{-3}}$ at $R= 50 {\rm km}$, which is the same as that used in NZv1. To determine $\langle \bar{f}_{ee}\rangle$, we introduce $\alpha$ defined as,
\begin{equation}
\alpha \equiv \frac{ n_{\bar{\nu}_e}  }{ n_{{\nu}_e}  }.
\label{eq:defalpha}
\end{equation}

$\barparena{\beta}_{ee}$ characterizes the shape of the angular distribution of neutrinos. This should be set in the range of $-2 \le \barparena{\beta}_{ee} \le 2$ so that $\barparena{f}_{ee}$ becomes positive in all angles (see Eq.~\ref{eq:iniang_in}). In this study we only consider the case with $\barparena{\beta}_{ee} \ge 0$, since the neutrinos where we consider in the situations for CCSN and BNSM have forward-peaked angular distributions outside of neutrino sphere. We note that$\barparena{\beta}_{ee} = 0$ corresponds to the case with a flat angular distribution in $\cos \theta_{\nu} \ge 0$, and that the degree of forward peaking increases with $\barparena{\beta}_{ee}$.

The existence of ELN crossing is a necessary and sufficient condition for occurrences of FFCs. We, hence, determine a set of parameters of $\alpha$ and $\barparena{\beta}_{ee}$ so that an ELN crossing appears in outgoing directions ($\cos \theta_{\nu} > 0$). From Eq.~\ref{eq:iniang_in}, the crossing point ($\cos \theta_{\nu (c)} $) can be analytically given as,
\begin{equation}
\cos \theta_{\nu (c)} = \frac{ (\alpha - 1) + 0.5( \beta_{ee} - \alpha \bar{\beta}_{ee}  ) }{  \beta_{ee} -  \alpha \bar{\beta}_{ee} }.
\label{eq:cross}
\end{equation}

\begin{figure}
   \includegraphics[width=\linewidth]{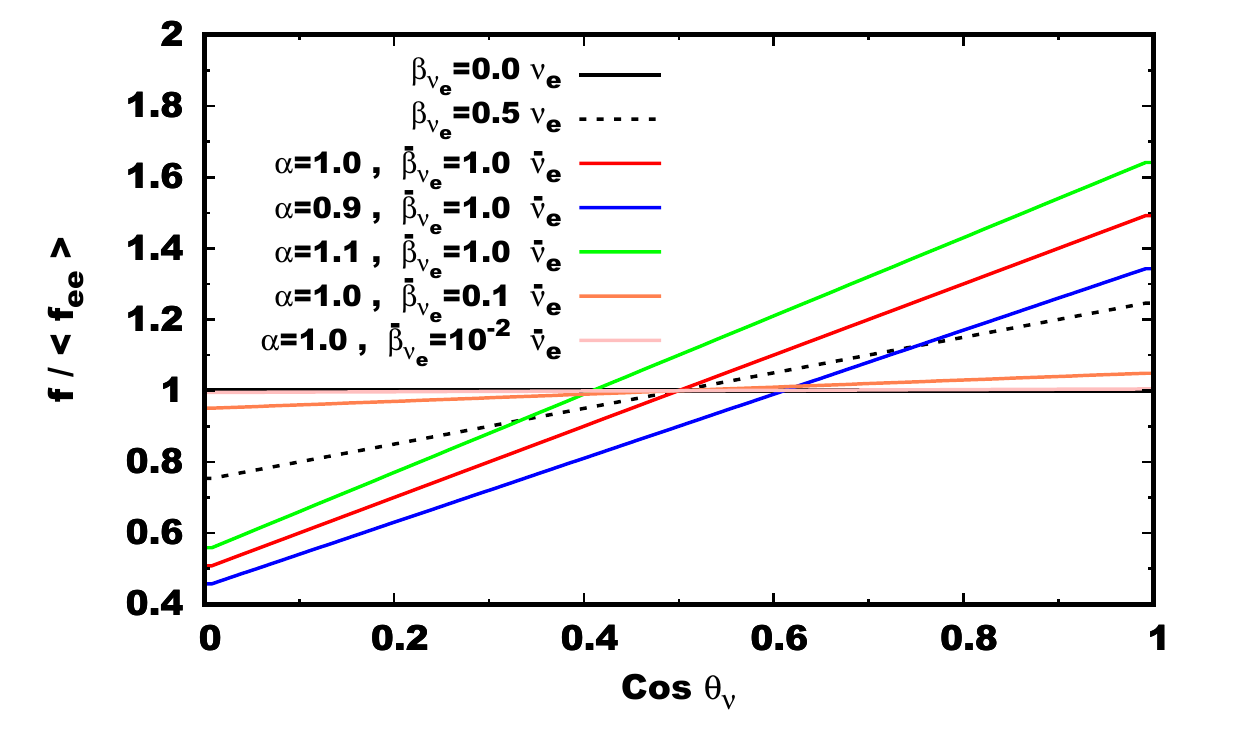}
   \caption{Angular distributions of $\nu_e$ and $\bar{\nu}_e$ at the inner boundary. We show some representative distributions used in our simulations. The vertical axis is normalized by $\langle f_{ee}\rangle$. See text for further details.
}
   \label{graph_IniAng}
\end{figure}

The set of parameters for all models is listed in Table~\ref{tab:model}. The angular distributions of $\nu_e$ and $\bar{\nu}_e$ at the inner boundary for some representative models are displayed in Fig.~\ref{graph_IniAng}. In our reference models (GL-Ref and LO-Ref), we set $\alpha=1$ (i.e., $\langle f_{ee}\rangle = \langle \bar{f}_{ee}\rangle$), $\beta_{ee}=0$, and $\bar{\beta}_{ee}=1$. We note that "GL" and "LO" in the name of these models denote "global" and "local" simulations, respectively; $\Delta R$ in Table~\ref{tab:model} represents the radial width of simulation box\footnote{The simulation box is $R_{in} \le r \le R_{in} + \Delta R$.}. It is worth to note that the neutrino angle of ELN crossing does not depend on $\barparena{\beta}_{ee}$ when we adopt $\alpha=1$ (see Eq.~\ref{eq:cross}); the crossing point is always located at $\cos \theta_{\nu (c)} = 0.5$. We also note that GL-Ref corresponds to the same initial condition that used in NZv1\footnote{It should be noted, however, that the numerical setup is not exactly the same as that used in NZv1. In NZv1, we changed the number density of neutrinos to control the degree of self-interaction potential, meanwhile the vacuum potential was not changed. In this study, we multiply $\xi$ for the total Hamiltonian (see Eq.~\ref{eq:BasEq}), implying that the vacuum potential is also affected by $\xi$.}. We run 17 QKE simulations in total with varying these parameters systematically.

The $\alpha-$dependence can be studied by comparing to the reference model to "GL(LO)-$\alpha$09" and "GL(LO)-$\alpha$11". In these models, the parameters are the same as those used in the reference model but for $\alpha=0.9$ and $1.1$, respectively. The ELN crossing point for "GL(LO)-$\alpha$09" and "GL(LO)-$\alpha$11" is $\cos \theta_{\nu (c)} = 11/18$ and $9/22$, respectively.

We study $\bar{\beta}_{ee}$-dependence under the choice of $\alpha=1$ and $\beta_{ee}=0$ (these are the same as those used in reference model). It is worth to note that $\bar{\beta}_{ee}$ characterizes the {\it depth} of ELN crossing. In fact, the crossing becomes shallower with decreasing $\bar{\beta}_{ee}$ (for instance, the depth becomes $\sim 1 \%$ for the choice of $\bar{\beta}_{ee}=10^{-2}$; see also Fig.~\ref{graph_IniAng}). We also note that the ELN crossing point does not depend on $\bar{\beta}_{ee}$ under the choice of $\alpha=1$. This suggests that we can study the sensitivity of FFC to the depth of ELN crossing under fixing the ELN crossing point. In this study, we consider four different cases: $\bar{\beta}_{ee}=1, 0.1, 10^{-2},$ and $10^{-3}$. In total, we run nine models (including reference models) to study the $\bar{\beta}_{ee}$ dependence. GL-$\bar{\beta}$01$\xi$-3 and LO-$\bar{\beta}$01 correspond to the case with $\bar{\beta}_{ee}=0.1$; GL-$\bar{\beta}$001$\xi$-2 and LO-$\bar{\beta}$001 are for the case with $\bar{\beta}_{ee}=10^{-2}$; GL-$\bar{\beta}$0001$\xi$-1, and LO-$\bar{\beta}$0001 represent the case for $\bar{\beta}_{ee}=10^{-3}$. We note that $\xi$ for the global models increases with decreasing $\bar{\beta}_{ee}$, i.e., relaxing the attenuation of Hamiltonian potential. This is possible because the depth of crossing becomes shallower with decreasing $\bar{\beta}_{ee}$, indicating that the growth rate and the oscillation wavelength of FFCs become slower and longer with decreasing $\bar{\beta}_{ee}$. On the other hand, we need to widen the size of simulation box for local simulations of these models with smaller values of $\bar{\beta}_{ee}$, since the default simulation box ($\Delta R = 10 {\rm m}$) would not be large enough to study the non-linear phase of flavor conversion. We, hence, set $\Delta R = 100 {\rm m}$ and $1 {\rm km}$ for LO-$\bar{\beta}$001 and LO-$\bar{\beta}$0001, respectively.

One of the striking results in this study is that strong flavor conversions can occur even in a very shallow ELN crossing such as the model with $\bar{\beta}_{ee}=10^{-3}$. To confirm that the result is not a numerical artifact, we carry out a resolution study (GL-H-$\bar{\beta}$0001$\xi$-1 model), in which we employ twice higher resolutions for both in space and neutrino angles than those of GL-$\bar{\beta}$0001$\xi$-1.

We also run simulations with a non-flat $\nu_e$ angular distribution (GL-$\beta$05$\xi$-3 and LO-$\beta$05). In these models, we adopt $\alpha=1$, $\beta_{ee}=0.5$, and $\bar{\beta}_{ee}=1$. For the sake of completeness, we also prepare "GL-Flip" and "LO-Flip" models. In these models, $\nu_e$ and $\bar{\nu}_e$ angular distributions are flipped from those used in the reference model. This corresponds to the case that $\bar{\nu}_e$ angular distributions are more forward-peaked than $\nu_e$. Although it may not be realistic in CCSN and BNSM environments, this model is meaningful to understand basic characteristics of FFCs.

Given the angular distributions at the inner boundary, we first run simulations with turning off neutrino oscillations, corresponding to classical neutrino transport, until the system reaches steady state. The obtained steady state distributions are used as initial conditions for QKE simulations. The radial and angular resolutions in our QKE simulations are summarized in Table~\ref{tab:model}. The resolution is set by reference to NZv1; $N_r=49152$ and $N_{\theta_{\nu}}=128$, the number of radial and neutrino angular grids, respectively, are sufficient to resolve flavor conversions and to capture qualitative trends in the non-linear phase. The physical time of global simulations is set to be $T_{\rm sim}=0.5 {\rm ms}$, which is a factor of $\sim 3$ longer than the light-crossing time of the simulation box (for neutrinos propagating along the radial direction). The physical time for local simulations are scaled by the ratio of spatial width of simulation box. We note that the local simulations in the present study cover more than ten times wider spatial region than those presented in our previous paper \cite{2022PhRvD.106f3011N}.

Before we move on to numerical results, three important caveats need to be mentioned. First, we assume in this study that incoming neutrinos ($\cos \theta_{\nu} < 0$) are dilute. However, they should be handled more precisely to study FFCs in CCSN and BNSM environments. As the radius decreases, the incoming neutrinos get more populated through neutrino emission and scattering with matter, and then neutrino angular distributions eventually become isotropic in optically thick regions. We also note that neutrinos undergo smooth transitions between optically thin and thick regions, suggesting that the discontinuous change of angular distributions at $\cos \theta_{\nu} = 0$ in our numerical setup is not realistic. This would affect FFC dynamics. In fact, the increase of neutrino number at $\cos \theta_{\nu} = 0$ may result in reducing the growth rate of FFCs (see, e.g., \cite{2022PhRvD.105l3003H}). It should be noted, however, that the effects of incoming neutrinos would be subdominant in the neutrino transparent region upon which we focus in this paper. We also note that angular distributions of incoming neutrinos hinge on the neutrino-matter interactions and spatial distributions of fluid. This exhibits that systematic studies are mandatory to study the impacts of incoming neutrinos on FFCs in CCSN and BNSM. Addressing this issue is beyond the scope of this paper.

Second, we inject neutrinos with ELN crossings from inner boundary in this study, but we need to keep in mind that this setup is artificial. In reality, ELN crossings are formed by interplay between neutrino advection and species-dependent matter interactions (see, e.g., \cite{2022arXiv220600676S,2022arXiv220704058S}), indicating that they appear smoothly with changing radius. 

Third, the interplay between advection and matter interactions also has an impact on the development of FFC itself. In fact, some recent studies have revealed that the enhancement or suppression of FFC by momentum-exchanged collisions hinge on neutrino angular distributions of neutrinos \cite{2022PhRvD.105l3003H}. Emission and absorption processes also cause decoherence, which typically work to suppress flavor conversion (but see collisional instability in \cite{2021arXiv210411369J,2022PhRvD.106j3029J,2022arXiv221009218L,2022arXiv221008254X,2022arXiv221203750X}). These processes would affect the overall properties of FFC including small scale structures. It should be mentioned, however, that local simulations do not have the ability to determine the actual impact of neutrino-matter interactions on flavor conversion due to the self-consistency issue \cite{2022PhRvD.106d3031J}, exhibits the necessity of large-scale simulations of FFCs with neutrino-matter interactions (see also \cite{2022arXiv220600676S,2022arXiv220704058S,2022arXiv221008254X}). We also note that, since the neutrino-matter interactions depend on fluid background, addressing this issue requires systematic studies of FFCs by varying fluid profile. This exhibits the same issue that we faced in the first caveat mentioned above. We, hence, leave these detailed studies of impacts of neutrino-matter interactions on FFCs in future work.

\begin{figure*}
   \includegraphics[width=\linewidth]{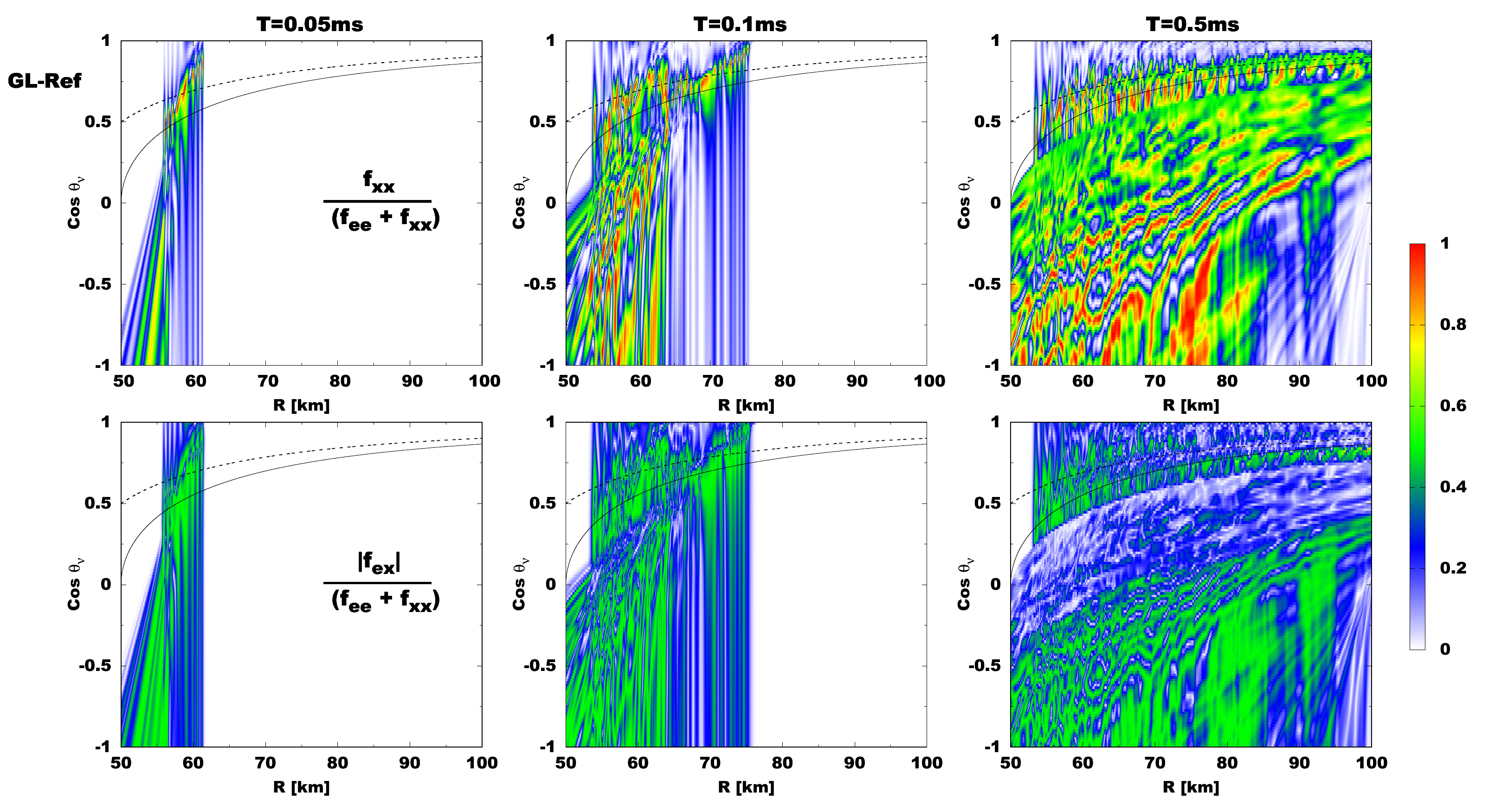}
   \caption{Top: color map of $f_{xx}$ as functions of radius and $\cos {\theta_{\nu}}$ for GL-Ref model. Bottom: the same as the left panel but for $|f_{ex}|$ (norm of off-diagonal component of density matrix of neutrinos). Both maps are normalized by $f_{ee}+f_{xx}$. From left to right, we display the result at ${\rm T} = 0.05, 0.1,$ and $0.5 {\rm ms}$, respectively. The solid line on each panel represents the angular trajectory for neutrinos emitted in the direction of $\cos {\theta_{\nu}}=0$ at the inner boundary ($R= 50 {\rm km}$). The dashed line portrays the angular trajectory for neutrinos emitted in the $\cos {\theta_{\nu}}=0.5$, which corresponds to the ELN crossing point for GL-Ref, at the inner boundary.
}
   \label{graph_Radi_vs_angular_GLrefNeut}
\end{figure*}

\begin{figure*}
   \includegraphics[width=\linewidth]{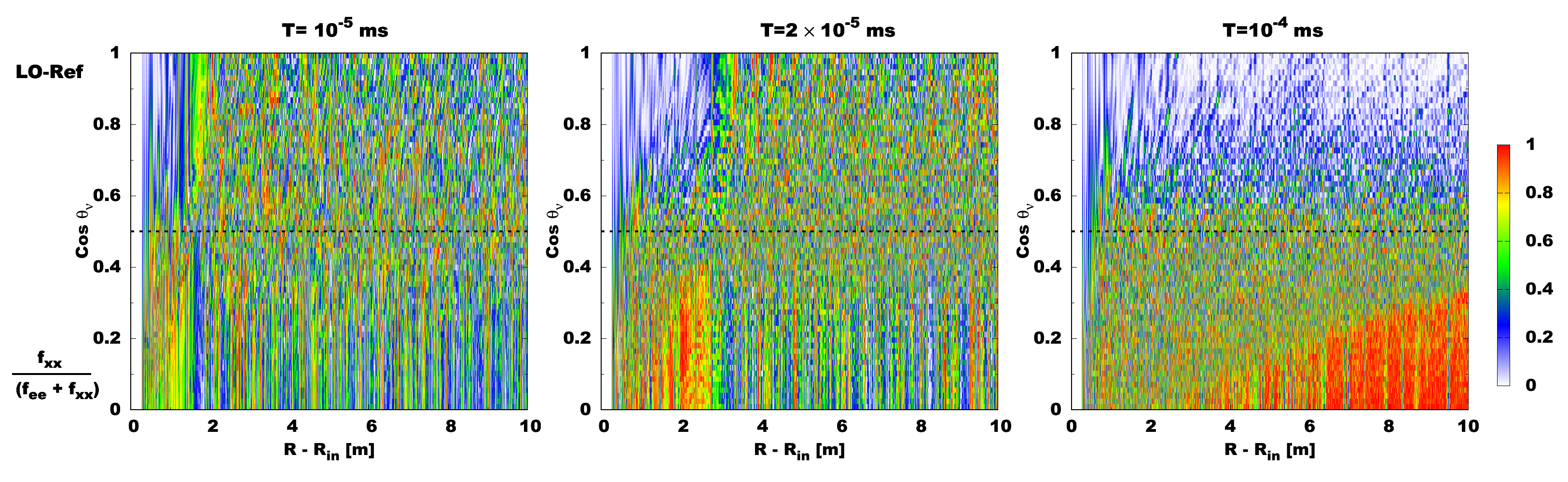}
   \caption{Color map of $f_{xx}/(f_{ee} + f_{xx})$ as functions of radius and neutrino angles for LO-Ref model. We focus on the angular distributions for outgoing neutrinos ($\cos {\theta_{\nu}} > 0$). The dashed line highlights $\cos {\theta_{\nu}} = 0.5$, which corresponds to the ELN crossing point for LO-Ref model.
}
   \label{graph_Radi_vs_angular_LOrefNeut}
\end{figure*}

\section{Result}\label{sec:result}
In this section, we present results of our numerical simulations. In most of the analysis, we focus on neutrinos, whereas we omit to display the relevant figures of antineutrinos, since they are almost identical to those of neutrinos.

\subsection{Dynamics}\label{subsec:dynamifeat}
We first focus on reference models (GL-Ref and LO-Ref) to show the overall trend of FFCs, and we postpone the detailed investigation of model dependence to Sec.~\ref{subsec:modeldepe}. Figure~\ref{graph_Radi_vs_angular_GLrefNeut} displays $f_{xx}$ and $|f_{ex}|$ (norm of off-diagonal component of density matrix) as functions of radius and neutrino angles at three different time snapshots ($T=0.05, 0.1,$ and $0.5 {\rm ms}$) for GL-Ref model. For visualization purposes, we normalize them by the value of $f_{ee}+f_{xx}$ at the same radius and neutrino angle in the color map. The solid and dashed lines represent the angular trajectory of neutrinos (as a function of radius) emitted in the direction of $\cos {\theta_{\nu}}= 0,$ and $0.5$, respectively, at the inner boundary ($R = 50 {\rm km}$). We note that $\cos {\theta_{\nu}}= 0.5$ corresponds to the neutrino angle where the ELN is zero, i.e., the zero-crossing point (see Sec.~\ref{sec:model}) at the inner boundary. These lines portray the transition of neutrino angular distributions to forward-peaked ones with increasing radius by a geometrical effect (spherical geometry).

The appearance of $\nu_x$ exhibits the occurrence of flavor conversion, since they are set to be zero initially and we do not inject them during the simulation (see also Sec.~\ref{sec:model}). The top panels of Fig.~\ref{graph_Radi_vs_angular_GLrefNeut} clearly show the appearance of $\nu_x$. On the other hand, $|f_{ex}|$ (see bottom panels of Fig.~\ref{graph_Radi_vs_angular_GLrefNeut}) represents the correlation between the two flavor states, which contains information on the vigor of flavor conversion. In the early phase (left panels), flavor conversions can be seen around the inner region of $\sim 60 {\rm km}$, and these neutrinos propagate outwards with time (see middle and right panels), that induces flavor conversions at large radii. We also find that the neutrino mixing can be mature enough to go through their non-linear saturation (left and middle panels of Fig.~\ref{graph_Radi_vs_angular_GLrefNeut}), and eventually the system reaches a quasi-steady state (right panels of Fig.~\ref{graph_Radi_vs_angular_GLrefNeut}).

Before entering into detailed discussions, there are three remarks that need to be mentioned. First, the flavor conversion observed in GL-Ref is dominated by fast mode, which is also confirmed by linear stability analysis (see Sec.~\ref{subsec:modeldepe}). The ratio of vacuum potential to self-interaction potentials is $\sim 10^{-8}$. Second, as displayed in all panels of Fig.~\ref{graph_Radi_vs_angular_GLrefNeut}, both $f_{xx}$ and $|f_{ex}|$ are very low around the inner (outer) boundaries for outgoing (incoming) neutrinos. This is attributed to the Dirichlet boundary condition (see Sec.~\ref{sec:model}). The injected $\nu_e$ and $\bar{\nu}_e$ from each boundary are constant in time, while other components of density matrix are set to be zero. This indicates that the spatial region near the boundary becomes the linearly growing regime of flavor conversions even after the system reaches quasi-steady phase. On the other hand, we attenuate the Hamiltonian potential in GL-Ref model ($\xi = 2 \times 10^{-4}$), leading to the artificial expansion of the spatial width for the linear growth regime. In fact, we found in NZv1 that this spatial width sensitively hinges on the choice of attenuation parameter (see Fig.~1 in NZv1). LO-Ref model suggests that the actual width (i.e., in the case with $\xi=1$) should be $\sim 20 {\rm cm}$ (see Fig.~\ref{graph_Radi_vs_angular_LOrefNeut}), which is also consistent with our previous study (see the left panel of Fig.~2 in \cite{2022PhRvD.106f3011N}).

Third, we find that FFCs in GL-Ref model arise at small radii, and then propagate outward with time. However, the propagation velocity observed in GL-Ref should be different from the case with $\xi=1$. In the case without attenuation of Hamiltonian potential, the growth of FFCs is so rapid, indicating that flavor conversions can be matured locally. This suggests that the spread of FFCs looks acausal, and such a feature is observed in LO-Ref. As shown in the left panel of Fig.~\ref{graph_Radi_vs_angular_LOrefNeut}, flavor conversion occurs in the entire simulation box at ${\rm T} = 10^{-5} {\rm ms}$, meanwhile the neutrinos emitted at inner boundary at ${\rm T}=0$ reaches up to $\sim 1 {\rm m}$. This suggests that the flavor conversion is not due to neutrino advection from the inner region but rather the local development of FFCs. It should be noted, however, that the self-interaction potential rapidly decreases with radius (since the neutrino number density decreases with radius and the forward-peaked angular distributions also causes to weaken the self-interaction potential), which indicates that the advection timescale eventually becomes shorter than that of FFCs at large radii.

One of the intriguing features displayed in Fig.~\ref{graph_Radi_vs_angular_GLrefNeut} is that flavor conversions for neutrinos emitted in the direction of $\cos \theta_{\nu} \gtrsim 0.5$ at the inner boundary are less vigorous than neutrinos emitted in other outgoing directions. In NZv1, the same feature was also observed. According to the convergence study for the attenuation of Hamiltonian performed in NZv1, this trend does not depend on the attenuation of Hamiltonian. In fact, the same feature of neutrino angular distribution also emerges in LO-Ref model. As shown in Fig.~\ref{graph_Radi_vs_angular_LOrefNeut}, the white color (i.e., less vigorous FFCs) in the region of $\cos \theta_{\nu} \ge 0.5$ spreads outwards roughly with a speed of light from inner to outer region. This exhibits that the inner boundary condition is responsible for the anisotropic angular distribution. It should be mentioned that local simulations performed in \cite{2022PhRvD.106f3011N} exhibited that the angular distributions of neutrinos are qualitatively different if we impose a periodic boundary condition (see Fig.~11 in the paper). We also note that more detailed investigations of effects of boundary conditions will be presented in another paper separately \cite{2022arXiv221109343Z}.

We also find that incoming neutrinos ($\cos \theta_{\nu} \le 0$) in the entire simulation box (but except for the vicinity of outer boundary) experience large flavor conversions. This is induced by outgoing neutrinos through the neutrino self-interactions. These neutrinos having non-zero $f_{ex}$ can advect inward from large radii, which potentially provide large seed perturbations to trigger FFCs at smaller radii. Since the vacuum contribution is suppressed by matter potential there, this can be a primary agent accelerating FFCs or even slow modes at semi-transparent and optically thick regions, which opens a new possibility to facilitate flavor conversions in CCSN and BNSM environments. Although quantifying the impact of inward advection of $f_{ex}$ on developments of FFCs is an important task, these studies require self-consistent treatments of neutrino-matter interactions. We leave the detailed study to future work.

\subsection{Temporal variations}\label{subsec:tempovari}

\begin{figure*}
   \includegraphics[width=\linewidth]{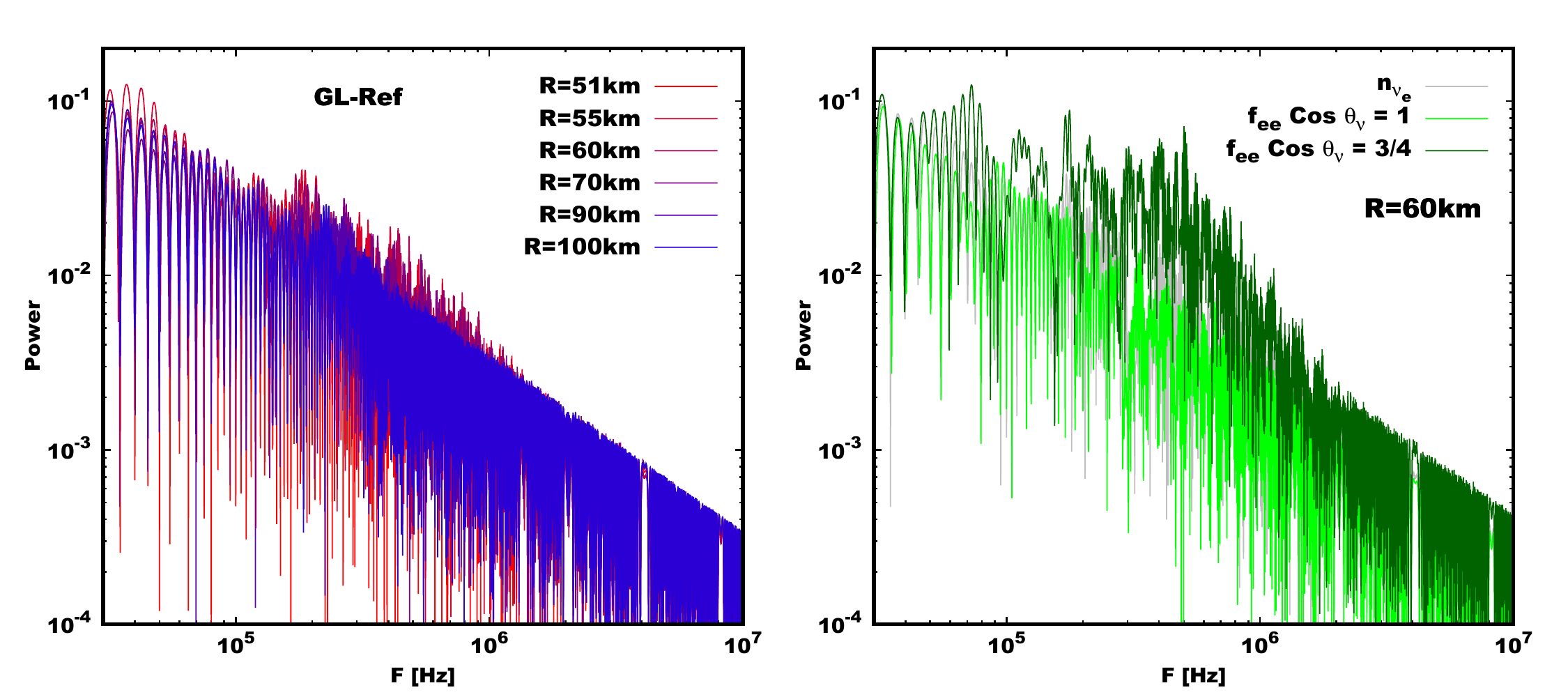}
   \caption{Finite-time Fourier transform of neutrinos for GL-Ref model. We employ Hann window function to compute the spectrum. We display the power spectrum normalized by that at $F=0$. Left: spectrum for $n_{\nu_e}$ (zeroth angular moment for $\nu_e$) at selected radii (specified by color). Right: spectrum for $f_{ee}$ for $\cos \theta_{\nu}=1$ and $3/4$ at $R = 60 {\rm km}$. For comparison, the result of $n_{\nu_e}$ at the same radius is also displayed as a gray line. We note that the Fourier analysis is performed by extending GL-Ref simulation; see text for more details.
}
   \label{graph_TFana_Gref}
\end{figure*}

\begin{figure}
   \includegraphics[width=\linewidth]{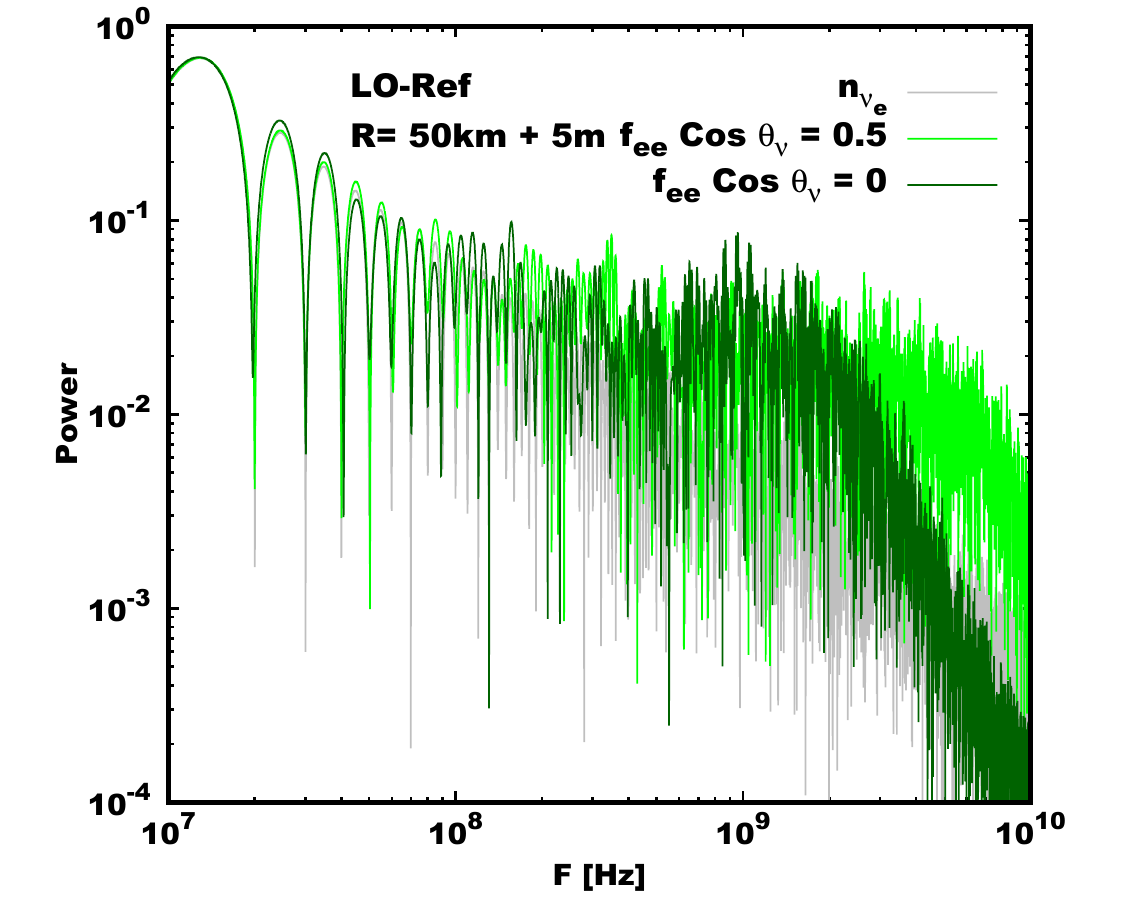}
   \caption{Same as the right panel of Fig.~\ref{graph_TFana_Gref} but for LO-Ref model. We compute the power spectrum at $R= 50 {\rm km} + 5 {\rm m}$.
}
   \label{graph_TFana_Lref}
\end{figure}

Temporal variations of neutrinos induced by FFCs are worth to be investigated. Fourier analysis is suited to capture their qualitative trend; hence, we compute them with particular focusing on the phase after the system reaches a quasi-steady state. In practice, however, the data output of our QKE simulations is not frequent enough to study the rapid oscillation of FFCs\footnote{As described in Sec.~\ref{sec:model}, we carry out simulations with high spatial resolutions, indicating that the high-cadence output presses a storage capacity. We output the data of GL-Ref simulation each $0.01 {\rm ms}$, which is $\sim 10$ times lower than required frequency. See text for more details.}, which prevents us from performing a post-processing analysis. We hence extend the simulation of GL-Ref from ${\rm T} = {\rm 0.5} {\rm ms}$ to ${\rm 1} {\rm ms}$, and we compute the Fourier integrals simultaneously during the extended simulation. This allows us to capture the rapid temporal variations of FFCs and it saves the storage.

Figure~\ref{graph_TFana_Gref} displays the Fourier transform of neutrinos at fixed spatial points for the extended GL-Ref model. In the computation of Fourier integral, we adopt the Hann window with the window size of $0.5 {\rm ms}$. The left panel of Fig.~\ref{graph_TFana_Gref} portrays the power spectrum for $n_{\nu_e}$ (zeroth angular moment for $\nu_e$) at different radii. We note that the vertical axis is normalized by the power at $F=0$, where $F$ denotes the frequency in Hertz. Although the temporal variation is rather mild for $n_{\nu_e}$, there are perceptible excess around the region of $10^{5} \lesssim F \lesssim 10^{6} {\rm Hz}$ at $R=55, 60, 70 {\rm km}$. On the other hand, the power around the same frequencies becomes weak at large radii (see e.g., blue lines). This is due to the phase cancellation of temporal variations. Neutrinos propagating different angles have different histories of FFCs, suggesting that the high frequency variation on each neutrino angle is random. Consequently, incoherent variations are canceled with each other through neutrino-self interactions, which damps the temporal variations.

The temporal variations of $f$ are different from those of the zeroth angular moment. $f$ in the angular region where FFCs occur vigorously varies with time even after FFCs reach non-linear saturation. This is one of the intrinsic features of collective neutrino oscillations (see, e.g., \cite{2015PhRvD..92l5030D,2015PhLB..751...43A}). The right panel of Fig.~\ref{graph_TFana_Gref} depicts the power spectrum for temporal variations of $f_{ee}$ with two different neutrino angles ($\cos \theta_{\nu}=1$ and $3/4$) at $R = 60 {\rm km}$. As clearly displayed in the panel, the temporal variation of $f$ strongly depends on the neutrino angle. $f_{ee}$ with $\cos \theta_{\nu}=1$ weakly varies with time, which is consistent with that FFCs are less vigorous in the angular region (see Fig.~\ref{graph_Radi_vs_angular_GLrefNeut}). On the other hand, $f_{ee}$ with $\cos \theta_{\nu}=3/4$ has strong temporal variations around the region of $10^{5} \lesssim F \lesssim 10^{6} {\rm Hz}$, and the power is remarkably higher than that of zeroth angular moment. This result suggests that the characteristic frequency of temporal variations are essentially common for all neutrinos, and they are also the same as their angular moments, whereas the random component of temporal variations in different neutrino angles is canceled when we compute the angular moment. This indicates that QKE solvers with lower angular moment schemes (such as two-moment methods) may not be capable o capturing intrinsic temporal variations of flavor conversion.

It should be mentioned that the temporal variation in GL-Ref is strongly affected by the attenuation of Hamiltonian potential ($\xi<1$). To quantify the impact, we performed the same Fourier analysis to LO-Ref as done to GL-Ref. We extended the simulation of LO-Ref from ${\rm T} = 1 \times 10^{-4} {\rm ms}$ to $2 \times 10^{-4} {\rm ms}$ with simultaneously computing Fourier integrals. The obtained power spectrum at $R= 50 {\rm km} + 5 {\rm m}$ is displayed in Fig.~\ref{graph_TFana_Lref}. As shown in the plot, we find some excess in the spectrum, and the characteristic frequency is $\sim 10^{4}$ times higher than GL-Ref, which is roughly consistent with $1/\xi \sim 5 \times 10^{3}$. We note that the slight difference of the frequency (a factor of $\sim 2$) is due to the difference of neutrino number density. The power spectrum of GL-Ref displayed in the right panel of Fig.~\ref{graph_TFana_Gref} is computed at $R= 60 {\rm km}$, where the neutrino number density becomes roughly a half of that at $R= 50 {\rm km} + 5 {\rm m}$. As a result, the characteristic frequency becomes a factor of two higher than $1/\xi$. Our analysis suggests that the temporal variation can be simply scaled by $1/\xi$ for the same neutrino number density. It should also be mentioned that the power spectrum of $f_{ee}$ for $\cos \theta_{\nu}=3/4$ is clearly higher than that of $n_{\nu_e}$ around the region of $10^{9} \lesssim F \lesssim 10^{10} {\rm Hz}$. This trend is common with that found in GL-Ref. We, hence, conclude that the attenuation of Hamiltonian potential does not compromise capturing the qualitative trend for the temporal variations of FFCs, although we need to multiply $F$ by a factor of $1/\xi$ to obtain the actual characteristic frequency.

\subsection{Model dependence}\label{subsec:modeldepe}

\begin{figure}
   \includegraphics[width=\linewidth]{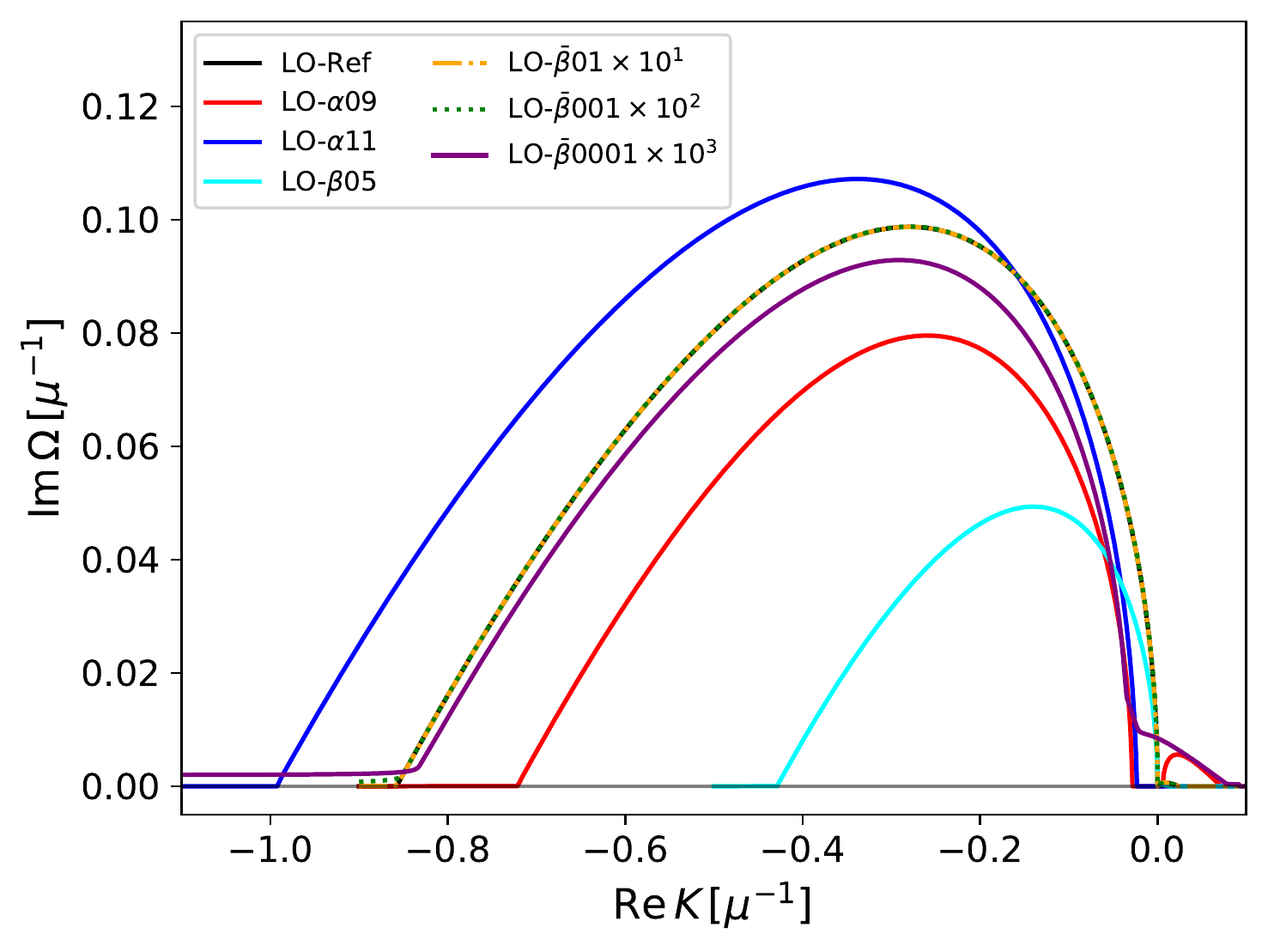}
   \caption{Dispersion relation (DR) of collective neutrino oscillations for models of local simulations ($\xi=1$, i.e., no attenuation of Hamiltonian potentials). In the computation of DR, the vacuum contribution is taken into account. The horizontal and vertical axes denote the wave number (${\rm Re} K$) and growth rate of their unstable modes (${\rm Im} \Omega$), respectively. In both axes, we show the result in unit of $\mu^{-1}$. In this plot, we omit to display the result of LO-Flip, which is the same as LO-Ref but changing the sign of ${\rm Re} K$. It should be noted that we multiply a factor of $1/\bar{\beta}_{ee}$ to both growth rate and wave number for LO-$\bar{\beta}$01, LO-$\bar{\beta}$001, and LO-$\bar{\beta}$0001 in this figure. This is not only for the visualization purpose but also for exhibiting that flavor conversion of LO-$\bar{\beta}$0001 is not dominated by fast mode; see text for more details.
}
   \label{DR_LO-scale}
\end{figure}

\begin{figure}
   \includegraphics[width=\linewidth]{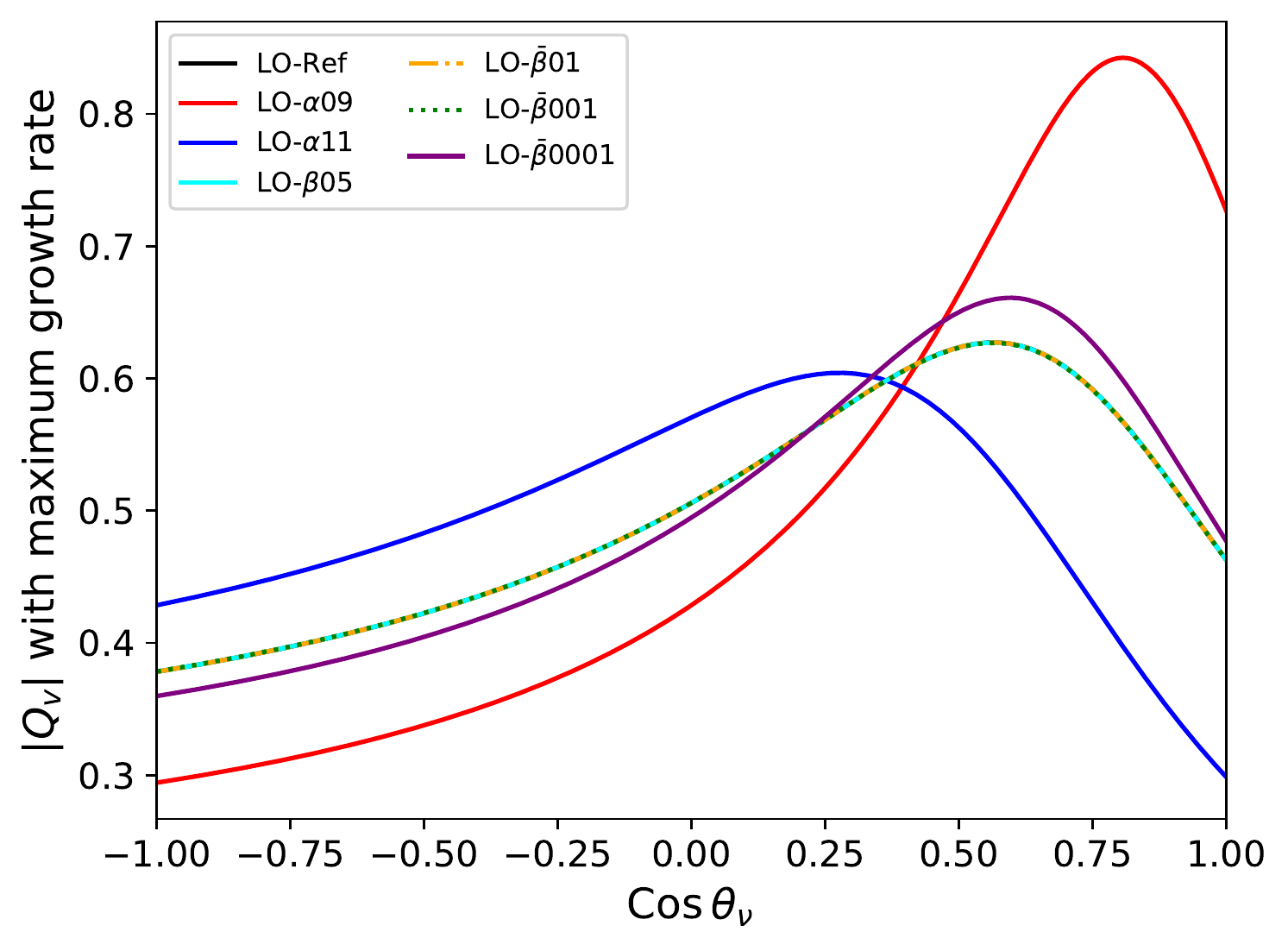}
   \caption{Eigenvectors ($Q_v$) with respect to the mode with maximum growth rate (see Fig.~\ref{DR_LO-scale}). They are normalized so as to be $\int_{-1}^{1} |Q_v| d \cos \theta_{\nu} = 1$.
}
   \label{Qv_LO}
\end{figure}

\begin{figure*}
   \includegraphics[width=\linewidth]{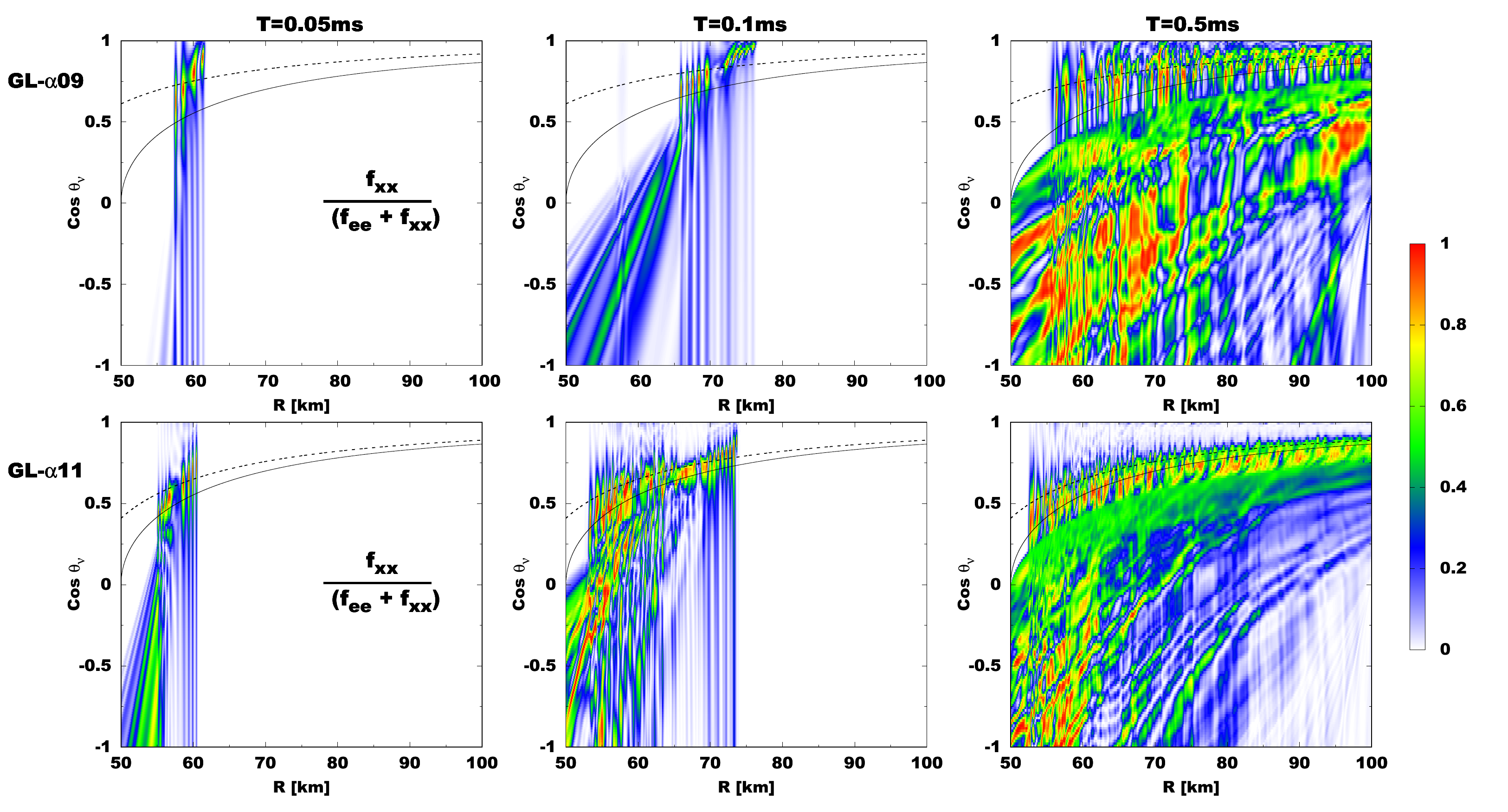}
   \caption{Same as the top panels of Fig.~\ref{graph_Radi_vs_angular_GLrefNeut} (color map of $f_{xx}/(f_{ee}+f_{xx})$ as functions of radius and $\cos {\theta_{\nu}}$) but for GL-$\alpha$09 (top) and GL-$\alpha$11 (bottom). We note that the ELN crossing angle in these models is different from that in GL-Ref; hence the dashed line on each panel is not identical to that displayed in Fig.~\ref{graph_Radi_vs_angular_GLrefNeut}.
}
   \label{graph_Radi_vs_angular_GLalphadepeNeut}
\end{figure*}

\begin{figure*}
   \includegraphics[width=\linewidth]{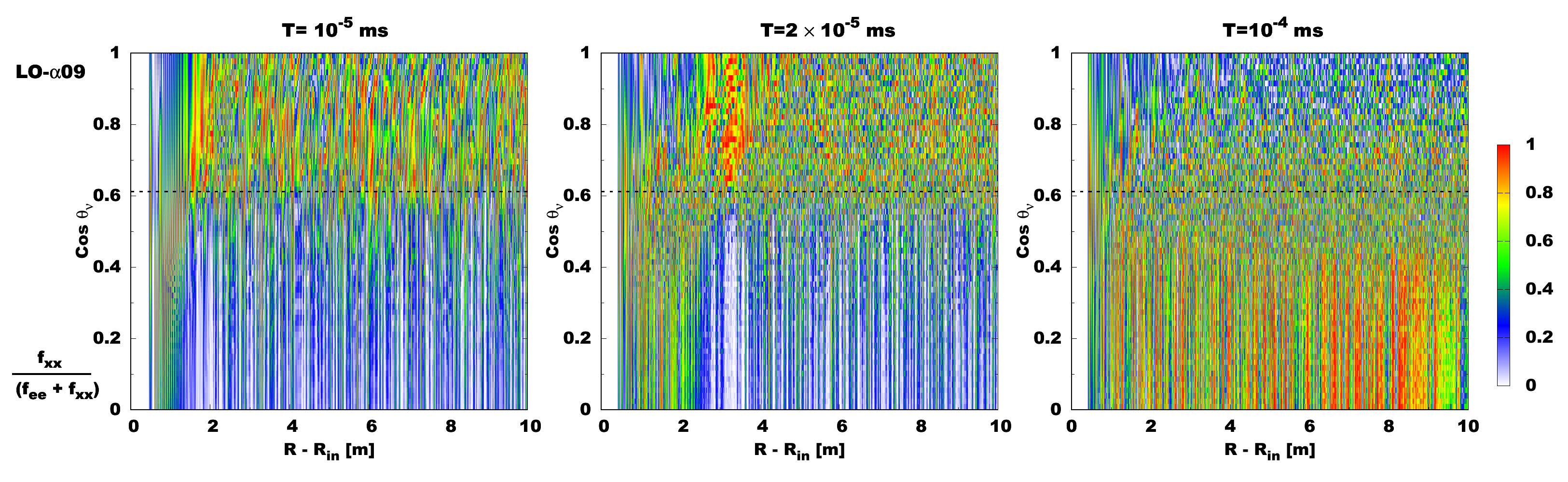}
   \caption{Same as Fig.~\ref{graph_Radi_vs_angular_LOrefNeut} but for LO-$\alpha$09 model. We note that the ELN crossing point is located at $11/18$, which is displayed with a dashed line in each panel.
}
   \label{graph_Radi_vs_angular_LOalpha09Neut}
\end{figure*}

\begin{figure*}
   \includegraphics[width=\linewidth]{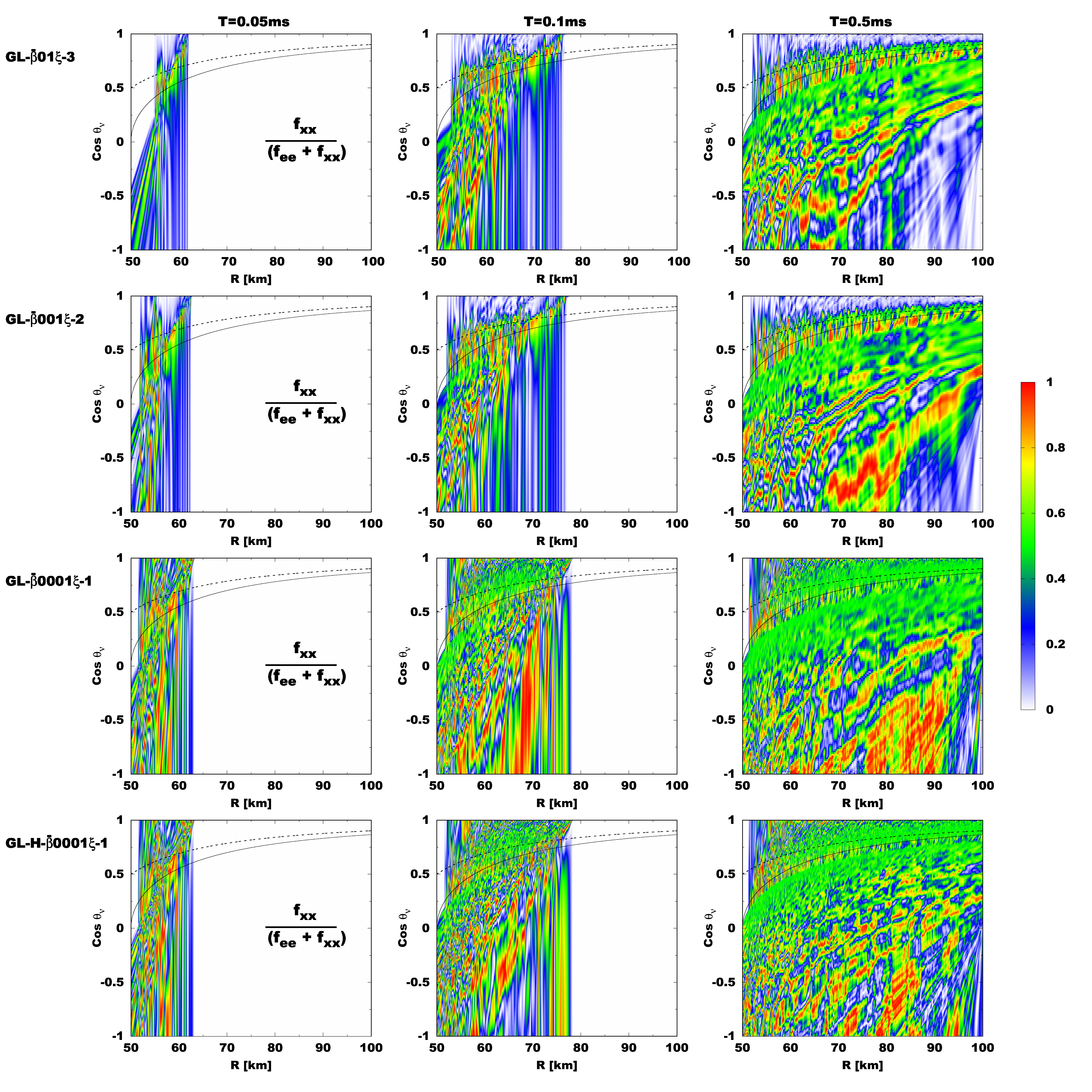}
   \caption{Same as the top panels of Fig.~\ref{graph_Radi_vs_angular_GLrefNeut} (color map of $f_{xx}/(f_{ee}+f_{xx})$ as functions of radius and $\cos {\theta_{\nu}}$) but for models with different $\bar{\beta}_{ee}$. From top to bottom, the panels display the result of GL-$\bar{\beta}$01$\xi$-3, GL-$\bar{\beta}$001$\xi$-2, GL-$\bar{\beta}$0001$\xi$-1, and GL-H-$\bar{\beta}$0001$\xi$-1, respectively.
}
   \label{graph_Radi_vs_angular_GLbarbetaeedepeNeut}
\end{figure*}

\begin{figure*}
   \includegraphics[width=\linewidth]{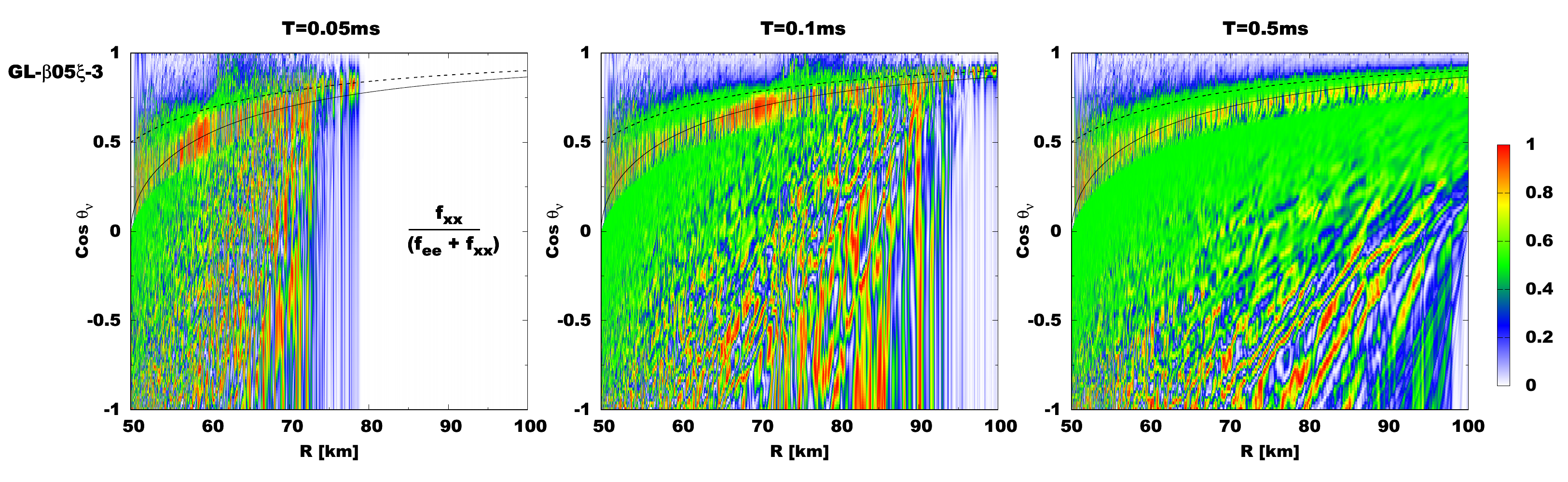}
   \caption{Same as the top panels of Fig.~\ref{graph_Radi_vs_angular_GLrefNeut} (color map of $f_{xx}/(f_{ee}+f_{xx})$ as functions of radius and $\cos {\theta_{\nu}}$) but for GL-$\beta$05$\xi$-3.
}
   \label{graph_Radi_vs_angular_GLbetaeedepeNeut}
\end{figure*}

\begin{figure*}
   \includegraphics[width=\linewidth]{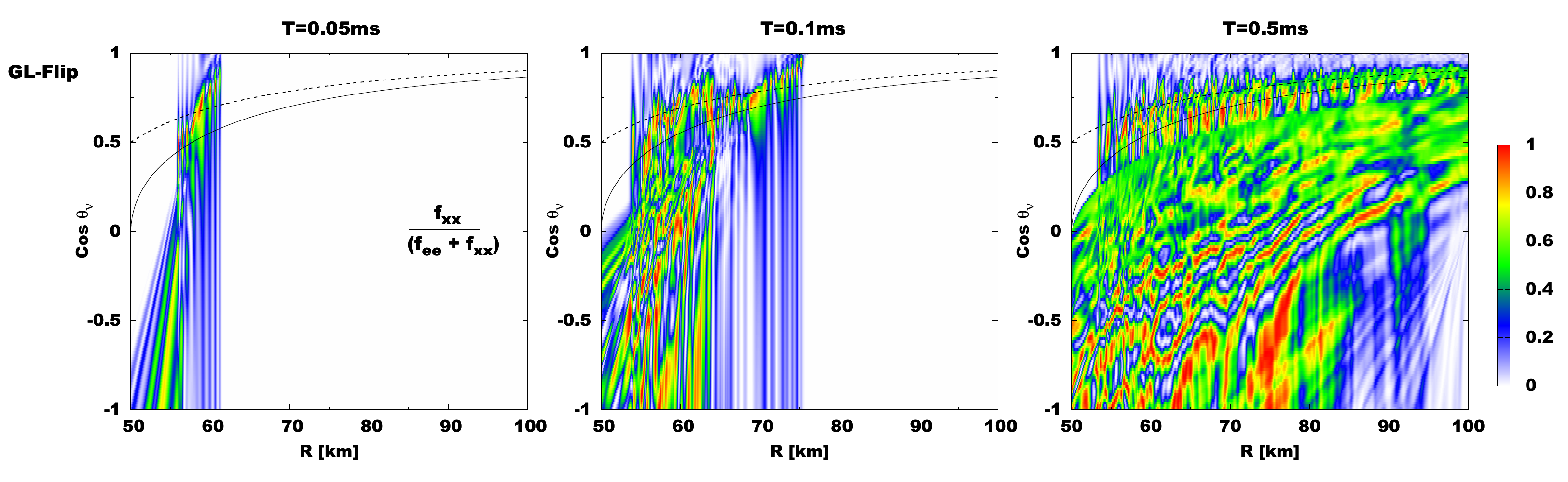}
   \caption{Same as the top panels of Fig.~\ref{graph_Radi_vs_angular_GLrefNeut} (color map of $f_{xx}/(f_{ee}+f_{xx})$ as functions of radius and $\cos {\theta_{\nu}}$) but for GL-Flip.
}
   \label{graph_Radi_vs_angular_GLFlipNeut}
\end{figure*}

Thus far, we have focused on GL-Ref and LO-Ref to discuss some basic properties of FFCs. We now turn our attention to model-dependent properties of FFCs. Before we begin the detailed discussion, we show some essential results of linear stability analysis of our models. Fig.~\ref{DR_LO-scale} portrays the dispersion relation (DR) of neutrinos at the inner boundary for local simulations (no attenuation of Hamiltonian potential, i.e., $\xi=1$). Both axes are normalized by $\mu^{-1}$, where $\mu \equiv \sqrt{2} G_F n_{\nu_e}$ ($G_F$ denotes the Fermi constant). In Fig.~\ref{Qv_LO}, we show the eigenvector ($Q_v$) for the mode with the maximum growth rate on each model. As shall be shown below, some intrinsic properties of flavor conversion can be extracted by combining the results of our QKE simulations with the DRs and eigenvectors.

Let us first discuss $\alpha$-dependent features of FFCs. Their overall features in global scales can be seen in Fig.~\ref{graph_Radi_vs_angular_GLalphadepeNeut}. As shown in the left and middle panels, the growth of flavor conversion in GL-$\alpha$11 (GL-$\alpha$09) is faster (slower) than GL-Ref. We note that the three models have the common $\xi (= 2 \times 10^{-4})$, indicating that $\xi$ does not affect the order of th growth rate. In fact, the growth feature of FFCs is consistent with the DR (see Fig.~\ref{DR_LO-scale}); the maximum growth rate for GL-$\alpha$11 (GL-$\alpha$09) model is higher (lower) than that in GL-Ref. At the end of these simulations (${\rm T} = 0.5 {\rm ms}$), we also find that the spatial region of linear growth regime for GL-$\alpha$11 (GL-$\alpha$09) model is narrower (wider) than GL-Ref (see right panels in Fig.~\ref{graph_Radi_vs_angular_GLalphadepeNeut}), which is also consistent with the DR. Let us make another remark; the ELN crossing angle in these two models is different from the reference model, which can be seen by comparing dashed lines in each panel to that in Fig.~\ref{graph_Radi_vs_angular_GLrefNeut}. The different ELN angular distribution changes the shape of DR (see Fig.~\ref{DR_LO-scale}), which gives an impact on FFCs in the non-linear phase (see below).

An interesting feature of FFCs emerges in non-linear regime. As shown in the right panels of Fig.~\ref{graph_Radi_vs_angular_GLalphadepeNeut}, FFCs occur in the almost entire neutrino angles for GL-$\alpha$09 (see top and right panel in Fig.~\ref{graph_Radi_vs_angular_GLalphadepeNeut}), whereas they appear vigorously only in the limited neutrino angles for GL-$\alpha$11. Consequently, the total amount of flavor conversion in GL-$\alpha$09 is higher than GL-$\alpha$11\footnote{The degree of flavor conversion will be quantified in Sec.~\ref{subsec:quasisteady}.}. This exhibits that the growth rate is not a good metric to determine the vigor of flavor conversion in the asymptotic state. We note that this is in line with the work of \cite{2022PhRvL.128l1102P}, which demonstrated by homogeneous simulations that the amount of flavor conversion does not always correlate with the growth rate.

We also find another intriguing feature in LO-$\alpha$09. As shown in Fig.~\ref{graph_Radi_vs_angular_LOalpha09Neut}, FFCs are matured earlier in the angular region of $\cos \theta_{\nu} \gtrsim \cos \theta_{\nu (c)}$. However, flavor conversions in other angles become more active with time, and eventually FFCs at $\cos \theta_{\nu} \lesssim \cos \theta_{\nu (c)}$ reach nearly flavor equipartition. This feature is a bit different from LO-Ref (see Fig.~\ref{graph_Radi_vs_angular_LOrefNeut}), in which flavor conversions grow rapidly in the all angles, and then those in $\cos \theta_{\nu} \gtrsim \cos \theta_{\nu (c)}$ become weaker after the neutrinos emitted from the inner boundary reaches there. The difference between the two models can be interpreted by linear stability analysis. As shown in Fig.~\ref{Qv_LO}, the eigenvector which has the maximum growth rate is very different from each other. In LO-$\alpha$09, the eigenvector has the sharper forward-peaking angular profile than that in LO-Ref, that accounts for the earlier development of FFCs in the region of $\cos \theta_{\nu} \gtrsim \cos \theta_{\nu (c)}$. On the other hand, one thing we do notice here is that the strong flavor conversion at $\cos \theta_{\nu} \lesssim \cos \theta_{\nu (c)}$ in the quasi-steady state is universal among models, which is a key feature to develop an approximate scheme of FFC (see Sec.~\ref{sec:phenomodel}).

Next, we consider the $\bar{\beta}_{ee}$-dependence of FFCs. As we have already mentioned in Sec.~\ref{sec:model}, $\bar{\beta}_{ee}$ represents the depth of ELN crossing; the smaller $\bar{\beta}_{ee}$ has the shallower crossing. As shown in Fig.~\ref{graph_Radi_vs_angular_GLbarbetaeedepeNeut}, we find that strong flavor conversion occurs even in very small $\bar{\beta}_{ee}$. One of the noticeable findings is that the angular distribution of neutrinos are almost identical between GL-Ref, GL-$\bar{\beta}$01$\xi$-3, and GL-$\bar{\beta}$001$\xi$-2. This property can be understood as follows. First, the shape of ELN angular distribution is identical among these models. More specifically, the only difference is the depth of ELN crossing. Under the assumption that vacuum potential can be neglected, FFCs properties should also be similar; the frequency of temporal variation is different but it can be scaled by the depth of ELN-crossing. This can be seen in Fig.~\ref{DR_LO-scale}, which exhibits that DRs of LO-$\bar{\beta}$01 and LO-$\bar{\beta}$001 become identical to that of DR when we multiply a factor of $1/\bar{\beta}_{ee}$ for both growth rate and the wave number. This is the reason why the dynamics and quasi-steady feature of GL-$\bar{\beta}$01$\xi$-3 and GL-$\bar{\beta}$001$\xi$-2 models is the same as GL-Ref.

Here, we make an important remark. It has been argued if large flavor conversions can occur in CCSNe, since the typical depth of ELN crossing may be an order of $\sim 1 \%$ according to some recent CCSN models (see, e.g., \cite{2019ApJ...886..139N,2022ApJ...924..109H}). However, our present result suggests that strong flavor conversion can happen even in such tiny ELN crossings. It should be stressed that more comprehensive study for realistic ELN angular distributions is needed to draw a more robust conclusion.

It should also be mentioned that GL-$\bar{\beta}$0001$\xi$-1 has qualitatively different properties from other models (see the third panels from top in Fig.~\ref{graph_Radi_vs_angular_GLbarbetaeedepeNeut}). For instance, the flavor conversion in the model occurs in the almost entire angles, whereas it is very weak around $\cos \theta_{\nu} = 1$ for other three models. The bottom panels in Fig.~\ref{graph_Radi_vs_angular_GLbarbetaeedepeNeut} exhibit that the result is not changed in the high resolution model, indicating that the difference is not due to numerical artifacts. The anomaly of the model can also be seen in the DR relation of LO-$\bar{\beta}$0001. As shown in Fig.~\ref{DR_LO-scale}, the scaled DR is clearly deviated from the reference model. This indicates that the vacuum potential affects the DR in GL-$\bar{\beta}$0001$\xi$-1, i.e., the maximum growth mode is not dominated by FFCs. This is understandable, since the depth of ELN crossing is tiny ($10^{-3}$). This illustrates that the contribution from the vacuum potential is no longer negligible in the model. The observed flavor conversion in GL-$\bar{\beta}$0001$\xi$-1 is, hence, affected by slow modes\footnote{We also note that the growth rate of LO-$\bar{\beta}$0001$\xi$ is smaller than that expected from fast mode (see Fig.~\ref{DR_LO-scale}). This is attributed to the fact that we adopt a positive $\Delta m^2$, i.e., normal-mass ordering, which works to suppress flavor conversions. We confirm that the trend becomes opposite, i.e., higher growth rate than the case only with FFCs for the case with inverted-mass hierarchy.}; consequently, the overall dynamics becomes different from other models.

In the models we have discussed so far, the $\nu_e$ angular distribution at the inner boundary is assumed to be flat in the region of $\cos {\theta_{\nu}} \ge 0$. In reality, however, $\nu_e$ has non-flat (forward-peaked) angular distributions, and therefore the dependence of FFCs upon $\nu_e$ angular distributions is also worth to be investigated. GL-$\beta$05$\xi$-3 (and LO-$\beta$05) provides an important information to this question. As shown in Fig.~\ref{graph_Radi_vs_angular_GLbetaeedepeNeut}, strong flavor conversions occur in GL-$\beta$05$\xi$-3, and their overall features are essentially the same as those in GL-Ref\footnote{In the simulation of GL-$\beta$05$\xi$-3, we employ $\xi = 2 \times 10^{-3}$, which is 10 times higher than that used in GL-Ref. This is possible because the linear growth rate of FFCs in LO-$\beta$05 is lower than GL-Ref (see Fig.~\ref{DR_LO-scale}).}. One thing we notice here is that the shape of ELN angular distribution is the same between GL-$\beta$05$\xi$-3 and GL-Ref, and the difference is the depth of the crossing. Since FFCs are dictated solely by ELN angular distributions, the overall dynamics for these models should be similar to each other. This is the reason why GL-$\beta$05$\xi$-3 has the similar dynamics as GL-Ref. As we shall quantify in Sec.~\ref{subsec:quasisteady}, however, the total amount of flavor conversion is not identical between the two models. This illustrates that the ELN angular distribution is not sufficient to determine the asymptotic states of $\nu_e$ (and $\bar{\nu}_e$), but rather we need species-dependent information.

Finally, we show the result of GL-Flip in Fig.~\ref{graph_Radi_vs_angular_GLFlipNeut}. As expected, the overall trend in GL-Flip is the same as that in GL-Ref. On the other hand, the total amount of flavor conversion of this model also slightly deviates from that in GL-Ref, which shall be quantified in the following section.

\subsection{Non-linear saturation and quasi-steady state}\label{subsec:quasisteady}

\begin{figure*}
   \includegraphics[width=\linewidth]{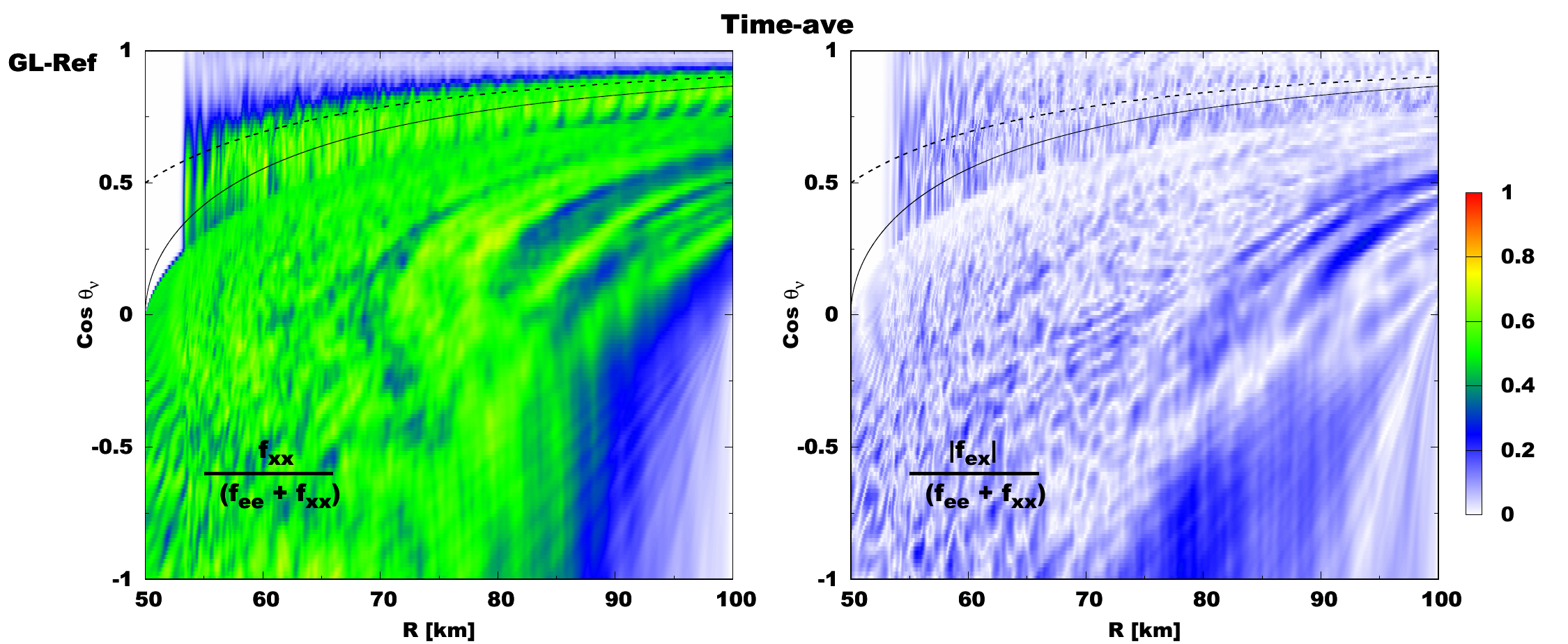}
   \caption{Color maps of time-averaged $f_{xx}/(f_{ee} + f_{xx})$ (left) and $|f_{ex}|/(f_{ee} + f_{xx})$ (right) as functions of radius and neutrino angles for GL-Ref model. The time-average is taken in the the time window of $0.3 {\rm ms} \le T \le 0.5 {\rm ms}$.
}
   \label{graph_Radi_vs_angular_GLrefNeut_Tave}
\end{figure*}

\begin{figure*}
   \includegraphics[width=\linewidth]{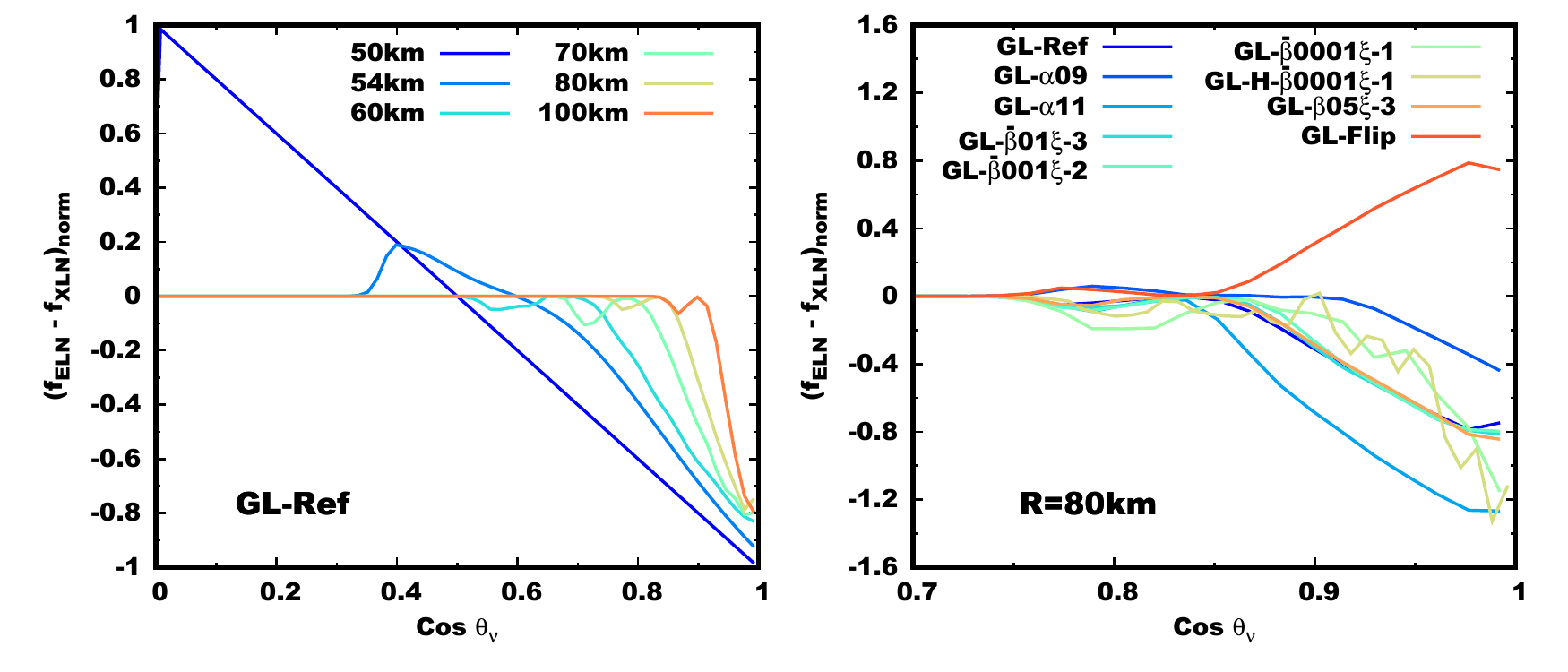}
   \caption{ELN-XLN angular distributions for $\cos {\theta_{\nu}} \ge 0$. In the left panel, we display the result at different radii for GL-Ref model. In the right panel, we compare ELN-XLN angular distributions for different models at $R = 80 {\rm km}$.
}
   \label{graph_ELNXLNang_Tave}
\end{figure*}

\begin{figure*}
   \includegraphics[width=\linewidth]{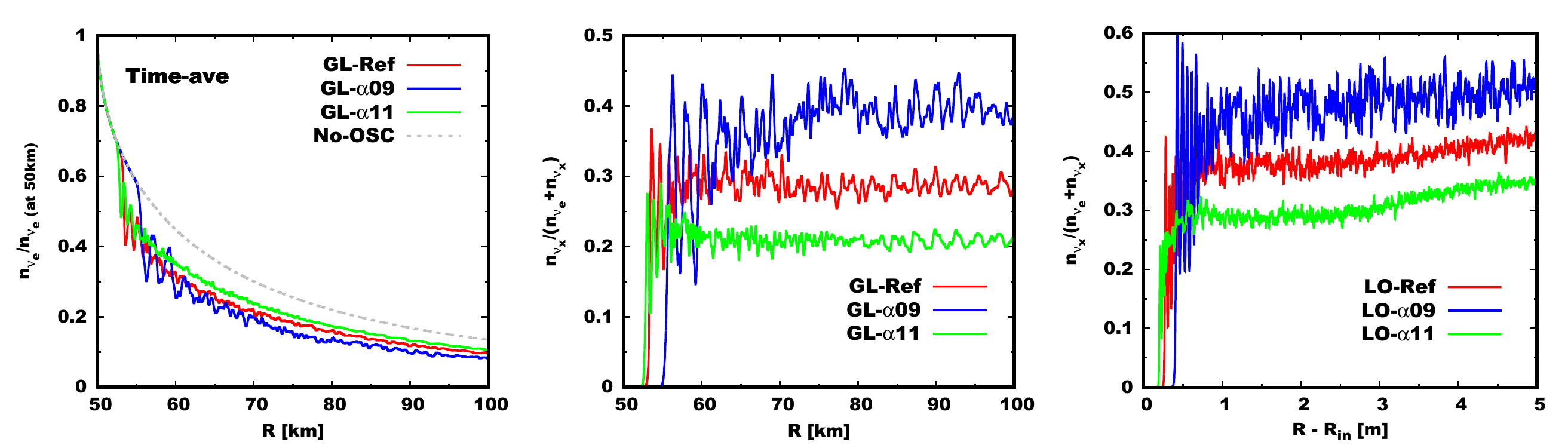}
   \caption{The time-averaged number density of neutrinos as a function of radius. We focus on $\alpha$-dependence in this figure. For global models (left and middle panels), the time-average is taken in the the time window of $0.3 {\rm ms} \le T \le 0.5 {\rm ms}$. For local simulations (the right panel), the window is $6 \times 10^{-5} {\rm ms} \le T \le 10^{-4} {\rm ms}$. In the left panel, we show the radial profile of $n_{\nu_e}$ normalized by that at the inner boundary. The middle panel displays $n_{\nu_x}/(n_{\nu_e} + n_{\nu_x})$, exhibiting the angular-averaged mixing degree of neutrinos for global simulations. The right panel shows the same as the middle panel but for local simulations.
}
   \label{graph_RadproNumdens_alphadepe}
\end{figure*}

\begin{figure*}
   \includegraphics[width=\linewidth]{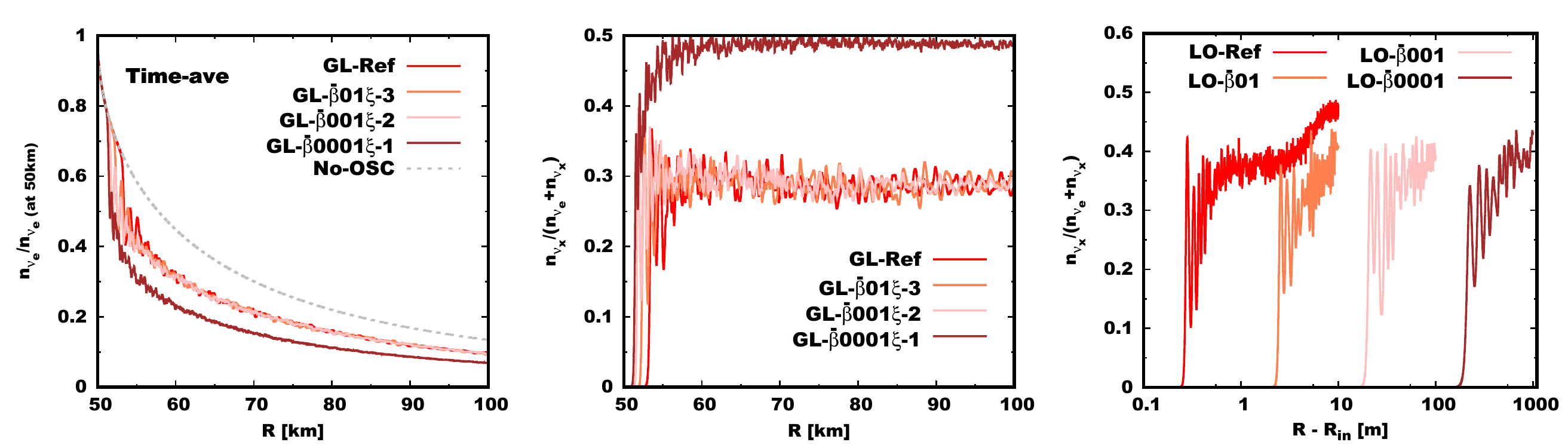}
   \caption{Same as Fig.~\ref{graph_RadproNumdens_alphadepe} but for the comparison of $\bar{\beta}_{ee}$ dependence. In the right panel, the time average for LO-$\bar{\beta}$01, LO-$\bar{\beta}$001, and LO-$\bar{\beta}$0001 is taken $6 \times 10^{-4} {\rm ms} \le T \le 10^{-3} {\rm ms}$, $6 \times 10^{-3} {\rm ms} \le T \le 10^{-2} {\rm ms}$, and $6 \times 10^{-2} {\rm ms} \le T \le 0.1 {\rm ms}$, respectively.
}
   \label{graph_RadproNumdens_betabardepe}
\end{figure*}

\begin{figure*}
   \includegraphics[width=\linewidth]{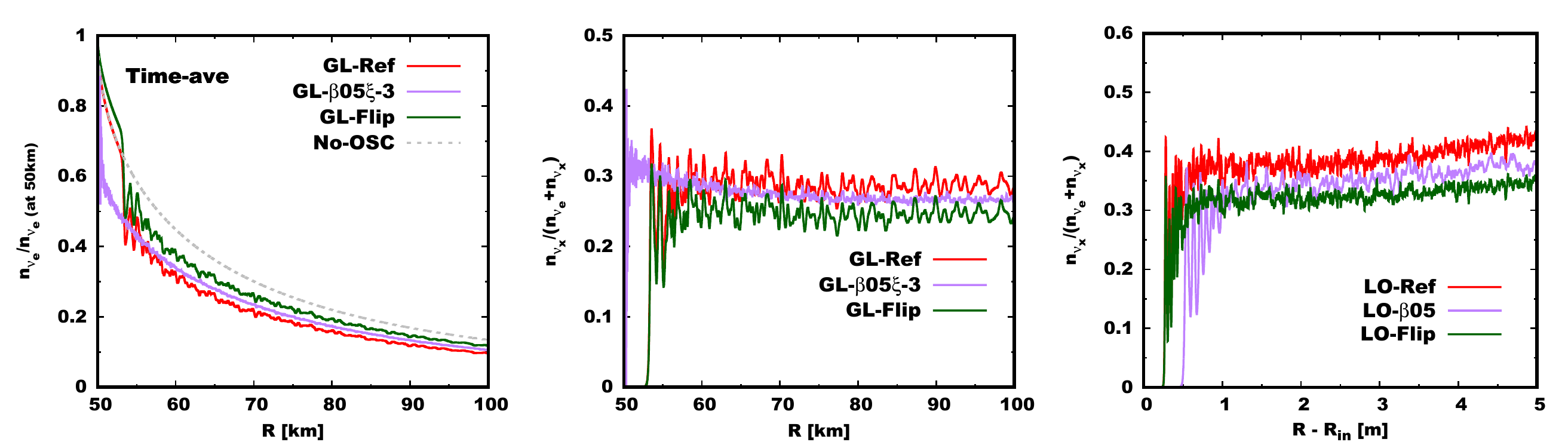}
   \caption{Same as Fig.~\ref{graph_RadproNumdens_alphadepe} but for comparing between Ref, $\beta$05$\xi$-3, and Flip models.
}
   \label{graph_RadproNumdens_betadepe_Flip}
\end{figure*}

In all models, FFCs undergo non-linear saturation and the system achieves a quasi-steady state. In this section, we underline their key properties, which provide important clues to develop approximate method for which to determine a quasi-steady state of FFCs without solving QKE (see Sec.~\ref{sec:phenomodel}).

In the left and right panels of Fig.~\ref{graph_Radi_vs_angular_GLrefNeut_Tave}, we display the time-averaged profile of $f_{ee}$ and $|f_{ex}|$ (the color map is normalized by $f_{ee} + f_{xx}$), respectively, as functions of radius and neutrino angles for GL-Ref. The time average is taken after the system establishes a quasi-steady state ($0.3 {\rm ms} \le t \le 0.5 {\rm ms}$). The left panel clearly exhibits that the time-averaged flavor conversion is very small for outgoing neutrinos in the angular region of $\cos {\theta_{\nu}} \gtrsim \cos \theta_{\nu (c)}$, whereas strong flavor conversions are observed in other angles. Another intriguing feature displayed in Fig.~\ref{graph_Radi_vs_angular_GLrefNeut_Tave} is that the time-averaged $|f_{ex}|$ is remarkably smaller than that of $f_{xx}$, and its typical value is $\lesssim 0.1$, exhibiting that the time-averaged neutrinos are essentially in flavor states. The angular dependence of $|f_{ex}|/(f_{ee} + f_{xx})$ is also weak compared to that of $f_{xx}$. Importantly, these trends are commonly observed in other models.

As discussed in NZv1, ELN-XLN angular distributions are useful quantities to characterize FFCs. In Fig.~\ref{graph_ELNXLNang_Tave}, we show these angular distributions measured at differet radii for GL-Ref in the left panel. In the figure, we focus on the angular region of $\cos {\theta_{\nu}} \ge 0$, and the vertical axis is normalized so that the norm of $f_{{\rm ELN}} - f_{{\rm XLN}}$ (where $f_{{\rm ELN}}$ and $f_{{\rm XLN}}$ are defined as $f_{ee} - \bar{f}_{ee}$ and $f_{xx} - \bar{f}_{xx}$, respectively) at $\cos {\theta_{\nu}} = 0$ becomes unity at the inner boundary. As shown in the left panel, the depth of ELN-XLN angular crossing decreases with radius. At $R = 54 {\rm km}$, the ELN-XLN crossing can be seen in the angular distribution, indicating that this is still in the linear growth regime for FFCs. At $R \sim 60 {\rm km}$, the angular crossing almost disappears. This trend is commonly observed in all large-scale simulations; see the right panel of Fig.~\ref{graph_ELNXLNang_Tave}. This panel portrays the ELN-XLN angular distribution of different models at $R=80 {\rm km}$. As can be seen in the figure, the ELN-XLN angular crossing is rather weak or disappears. We note that the ELN-XLN angular distributions in GL-Flip and GL-Ref have a mirror symmetry with respect to $f_{{\rm ELN}} - f_{{\rm XLN}}=0$. This is due to the fact that the angular distributions of $\nu_e$ and $\bar{\nu}_e$ in GL-Flip are swapped from those in GL-Ref model.

Our result suggests that the disappearance of ELN-XLN angular crossing in the time-averaged profile is one of the common properties of FFCs. This can be understood through the linear stability analysis. As long as the off-diagonal component is remarkably smaller than diagonal ones (this condition is actually satisfied in our simulations; see the right panel of Fig.~\ref{graph_Radi_vs_angular_GLrefNeut_Tave}), the linear analysis provides a reasonable diagnostics for the stability, and it suggests that the existence of ELN-XLN angular crossing provides a necessary and sufficient condition for instability of fast mode.

Below, we discuss the total amount of neutrino-flavor conversion in a quasi-steady state. From Figs~\ref{graph_RadproNumdens_alphadepe}~to~\ref{graph_RadproNumdens_betadepe_Flip}, we show the time-averaged number density of neutrinos as a function of radius. In Fig.~\ref{graph_RadproNumdens_alphadepe}, we focus on $\alpha$-dependence. As shown in the left panel, $n_{\nu_e}$ decreases almost discontinuously at $R \sim 55 {\rm km}$, exhibiting that FFCs enters into the non-linear phase. It is important to note that the change of $n_{\nu_e}$ in GL-$\alpha$09 is the largest among the three models, which is also consistent with the middle panel, in which we show the number density of $n_{\nu_x}$ normalized by $n_{\nu_e} + n_{\nu_x}$. These panels clearly show that the total amount of neutrino-flavor conversion in GL-$\alpha$09 is the highest among the three models. In local simulations, we also find that the same trend, which can be seen in the right panel of Fig.~\ref{graph_RadproNumdens_alphadepe}\footnote{We display the radial profile of $n_{\nu_x}$ normalized by $n_{\nu_e} + n_{\nu_x}$ only for the spatial region of $R_{in} \le R \le R_{in} + 5 {\rm m}$ in the right panel of Fig.~\ref{graph_RadproNumdens_alphadepe}. We note that the neutrinos at larger radii in local simulations have not reached a quasi-steady state. This is because the neutrinos propagating in the direction of $\cos{\theta_{\nu}} \sim 0$ stagnates around the initial position. This suggests that we need much longer time simulations than the light-crossing time of neutrinos with $\cos{\theta_{\nu}} =1$ so that all neutrinos interact with each other in the computational domain. On the other hand, it is not our purpose that we make the whole system of local simulations establish the quasi-steady state. Rather, we use local simulations to interpret the result of large-scale ones.}. On the other hand, as we discussed in Sec.~\ref{subsec:dynamifeat}, the growth rate of flavor conversion in LO-$\alpha$09 is the lowest among the three models. This illustrates that the growth rate and saturation amplitude do not correlate with each other.

In Fig.~\ref{graph_RadproNumdens_betabardepe}, we display the same quantities as those shown in Fig.~\ref{graph_RadproNumdens_alphadepe} but focusing on $\bar{\beta}_{ee}$-dependence. It should be stressed again that strong flavor conversions occur even in the shallow ELN-crossing at the inner boundary, and the saturation of flavor mixing is almost universal among GL-Ref, GL-$\bar{\beta}$01$\xi$-3, and GL-$\bar{\beta}$001$\xi$-2. We also note that flavor conversion in GL-$\bar{\beta}$0001$\xi$-1 is not dominated by fast modes; consequently, the saturation property deviates from others (see also Sec.~\ref{subsec:dynamifeat}). In fact, the angular structure of flavor conversion in GL-$\bar{\beta}$0001$\xi$-1 is remarkably different from GL-Ref (see the middle panel in Fig.~\ref{graph_RadproNumdens_betabardepe}), and the total amount of flavor conversion becomes the highest among these models, despite of the fact that the growth rate is much smaller than that of GL-Ref (see in the right panel of Fig.~\ref{graph_RadproNumdens_betabardepe}). This is also consistent with the above argument that the growth rate does not determine the total amount of flavor conversion.

In Fig.~\ref{graph_RadproNumdens_betadepe_Flip}, we display the results for GL-Ref, GL-$\beta$05$\xi$-3, and GL-Flip. It should be mentioned that the shape of ELN angular distribution at the inner boundary is common among all models. The difference from GL-Ref is the depth of ELN crossing and the sign of self-interaction potential for GL-$\beta$05$\xi$-3 and GL-Flip, respectively. Although the total amount of flavor conversion is almost identical among these three models, we find that they are not exactly identical. The flavor conversion in GL-$\beta$05$\xi$-3 is slightly lower than GL-Ref, and GL-Flip is the lowest among these three models. The same trend can be observed in the local simulations (see the right panel in Fig.~\ref{graph_RadproNumdens_betadepe_Flip}).

This result is also another evidence that the total amount of flavor conversion can not be determined by ELN and XLN distribtuions but rather we need species-depedent information. Our interpretation for this argument is as follows. If there are many neutrinos and antineutrinos in the angular region where FFCs occur vigorously, the total amount of flavor conversion also becomes large. It is important to note that the number of both neutrinos and antineutrinos can be increased at specific angular directions with sustaining the ELN-XLN angular distributions\footnote{This is because the ELN (XLN) represents the {\it difference} between electron(heavy leptonic)-type neutrinos and their antineutrinos.}. This is a key to understand the trend observed in our simulations. We find that the total amount of flavor conversion tends to be large for models that have large number of neutrinos in the angular region where flavor conversions vigorously occur. This also exhibits a possibility that large flavor conversions can be induced even if they occur in narrow angular regions, since we can centralize neutrinos in the unstable angular region without changing ELN and XLN distributions.

In the following section, we provide a new approximate method for which to determine quasi-steady state of neutrino distributions in FFCs. Our proposed approach captures some key trends of FFCs that we have discussed above. Our method provides a useful way to incorporate effects of FFCs in classical neutrino transport methods including full Boltzmann neutrino transport and other approximate methods used in CCSN and BNSM simulations.

\section{Approximate scheme to determine quasi-steady state of FFC}\label{sec:phenomodel}

\subsection{Basis}\label{subsec:basis}

Before we begin, let us make some remarks. First, the proposed method should be considered provisional. Although they are in reasonable agreement with FFC simulations presented in this paper, more systematic studies are required to assess if our method can capture all key features of quasi-steady state for arbitrary FFCs. One of the major concerns is the applicability to FFCs in semi-transparent and optically thick regions, since the background neutrino angular distributions are qualitatively different from those studied in this paper. We also note that interplay between flavor conversions and neutrino-matter interactions would affect the asymptotic state of FFCs. Another limitation in our proposed method is that it is developed based on the assumption that initial angular distributions of neutrinos have single ELN-XLN crossings. Although single crossings would be the majority in the optically thin region \cite{2019ApJ...886..139N,2022ApJ...924..109H}, multiple crossings would occur in the vicinity of PNS \cite{2020PhRvD.101b3018D,2020PhRvD.101f3001G}. 

It is interesting to compare our approximate scheme to others (see, e.g., \cite{2022arXiv220505129B}). It should be mentioned, however, that other approximate methods are developed based on local simulations with a periodic boundary condition, which is different from ours. Since the boundary condition has strong influence on angular structure of FFCs in quasi-steady states \cite{2022PhRvD.106f3011N}, our approximate scheme would not be compatible with others. We leave a more in-depth analysis for impacts of boundary conditions on quasi-steady states to another paper \cite{2022arXiv221109343Z}.

The key idea of our method is that we determine neutrino distributions in quas-steady states so as to eliminate ELN-XLN angular crossings (see Sec.~\ref{subsec:quasisteady}). Another important indication from our numerical simulations is that FFCs are always vigorous in the region of $\cos \theta_{\nu} \lesssim \cos \theta_{\nu (c)}$, where $\theta_{\nu (c)}$ denotes the ELN-XLN angular crossing for the initial distribution of neutrinos. Interestingly, it does not depend on angular structure of eigenvectors of unstable modes obtained by linear stability analysis (see Fig.~\ref{Qv_LO}). In fact, FFCs are strong in the region of $\cos \theta_{\nu} \lesssim \cos \theta_{\nu (c)}$ after the system reaches a quasi-steady state, despite of the fact that $|Q_v|$ at $\cos{\theta_{\nu}} \sim 1$ is remarkably higher than those at $\cos{\theta_{\nu}} \sim 0$ (for instance, see $|Q_v|$ for LO-$\alpha$09 in Fig.~\ref{Qv_LO}).

We make two remarks on our claim that FFCs are always strong in the angular region of $\cos \theta_{\nu} \lesssim \cos \theta_{\nu (c)}$. First, the vigor of FFCs in the region of $\cos \theta_{\nu} \lesssim \cos \theta_{\nu (c)}$ would be due to effects of radial advection. The radial velocity of neutrinos is proportional to $\cos \theta_{\nu}$, indicating that the neutrinos in the region of $0 < \cos \theta_{\nu} \lesssim \cos \theta_{\nu (c)}$ slowly propagate in the outgoing radial direction. This suggests that those neutrinos have enough time to grow FFCs before they advect. Second, our result can be applied only for the forward-peaked angular distributions of neutrinos. More specifically, the condition can be written as,
\begin{equation}
  \begin{split}
\int_{-1}^{0} |f_{\rm ELN} - f_{\rm XLN}|  d \cos{\theta_{\nu}} < \int_{0}^{1} |f_{\rm ELN} - f_{\rm XLN}| d \cos{\theta_{\nu}}.
  \end{split}
\label{eq:ELNXLNoutindomi}
\end{equation}
If this inequality is not satisfied, our proposed method would not be a good approximation. Although this is an issue which we need to improve, the proposed method is still powerful for CCSN and BNSM, since the condition of Eq.~\ref{eq:ELNXLNoutindomi} is always satisfied in the optically thin region upon which we are currently focusing in this paper.

In our approximate scheme, we determine the angular-dependent survival-probability of neutrinos from initial states to quasi-steady ones. In the two-flavor approximation, the asymptotic state of neutrinos ($f^{\rm 2fl}$) can be written in terms of survival-probability of neutrinos ($p^{\rm 2fl}$) and antineutrinos ($\bar{p}^{\rm 2fl}$) as,
\begin{equation}
 \begin{split}
&f^{\rm 2fl}_{e} (\theta_{\nu}) = p^{\rm 2fl} (\theta_{\nu}) f^{0}_{e} (\theta_{\nu}) + \left(1- p^{\rm 2fl} (\theta_{\nu}) \right) f^{0}_{x} (\theta_{\nu}), \\
&\bar{f}^{\rm 2fl}_{e} (\theta_{\nu}) = \bar{p}^{\rm 2fl} (\theta_{\nu}) \bar{f}^{0}_{e} (\theta_{\nu}) + \left(1- \bar{p}^{\rm 2fl} (\theta_{\nu}) \right) \bar{f}^{0}_{x} (\theta_{\nu}), \\
&f^{\rm 2fl}_{x} (\theta_{\nu}) =  \left(1- p^{\rm 2fl} (\theta_{\nu}) \right) f^{0}_{e} (\theta_{\nu}) + p^{\rm 2fl} (\theta_{\nu})  f^{0}_{x} (\theta_{\nu}), \\
&\bar{f}^{\rm 2fl}_{x} (\theta_{\nu}) =  \left(1- \bar{p}^{\rm 2fl} (\theta_{\nu}) \right) \bar{f}^{0}_{e} (\theta_{\nu}) + \bar{p}^{\rm 2fl} (\theta_{\nu}) \bar{f}^{0}_{x} (\theta_{\nu}),
 \end{split}
\label{eq:trapro_two}
\end{equation}
where $f^{0}$ denotes the initial state of neutrinos. In the three-flavor case, it can be written as \cite{2000PhRvD..62c3007D,2021MNRAS.500..696N,2021MNRAS.500..319N},
\begin{equation}
 \begin{split}
&f^{\rm 3fl}_{e} (\theta_{\nu}) = p^{\rm 3fl} (\theta_{\nu}) f^{0}_{e} (\theta_{\nu}) + \left(1- p^{\rm 3fl} (\theta_{\nu}) \right) f^{0}_{x} (\theta_{\nu}), \\
&\bar{f}^{\rm 3fl}_{e} (\theta_{\nu}) = \bar{p}^{\rm 3fl} (\theta_{\nu}) \bar{f}^{0}_{e} (\theta_{\nu}) + \left(1- \bar{p}^{\rm 3fl} (\theta_{\nu}) \right) \bar{f}^{0}_{x} (\theta_{\nu}), \\
&f^{\rm 3fl}_{x} (\theta_{\nu}) = \frac{1}{2}\left(1- p^{\rm 3fl} (\theta_{\nu}) \right) f^{0}_{e} (\theta_{\nu}) + \frac{1}{2} \left(1+p^{\rm 3fl} (\theta_{\nu}) \right)  f^{0}_{x} (\theta_{\nu}) , \\
&\bar{f}^{\rm 3fl}_{x} (\theta_{\nu}) = \frac{1}{2} \left(1- \bar{p}^{\rm 3fl} (\theta_{\nu}) \right) \bar{f}^{0}_{e} (\theta_{\nu}) + \frac{1}{2} \left(1+\bar{p}^{\rm 3fl}  (\theta_{\nu}) \right) \bar{f}^{0}_{x} (\theta_{\nu}) ,
 \end{split}
\label{eq:trapro_three}
\end{equation}
where we assume $f^{0}_{x} = f^{0}_{\mu} = f^{0}_{\tau}$ and $\bar{f}^{0}_{x} = \bar{f}^{0}_{\mu} = \bar{f}^{0}_{\tau}$, which are reasonable conditions for neutrinos in CCSN and BNSM. The flavor equipartition can be obtained by $p^{\rm 2fl}=\bar{p}^{\rm 2fl}=1/2$ and $p^{\rm 3fl}=\bar{p}^{\rm 3fl}=1/3$ for two-flavor and three-flavor cases, respectively.

There are mainly two noticeable properties of FFCs. First, the survival-probability does not depend on neutrino-energy; hence, we drop the energy-dependence in Eqs.~\ref{eq:trapro_two}~and~\ref{eq:trapro_three}. Second, FFCs induce pairwise neutrino-flavor conversions, i.e., $p(\theta_{\nu})=\bar{p}(\theta_{\nu})$. It should be mentioned, however, that the angular-averaged survival-probability for neutrinos and antineutrinos are, in general, different from each other, which is simply because $f^{0}$ is not equal to $\bar{f}^{0}$.

\subsection{Implementation}\label{subsec:FFCM}

\begin{figure*}
   \includegraphics[width=\linewidth]{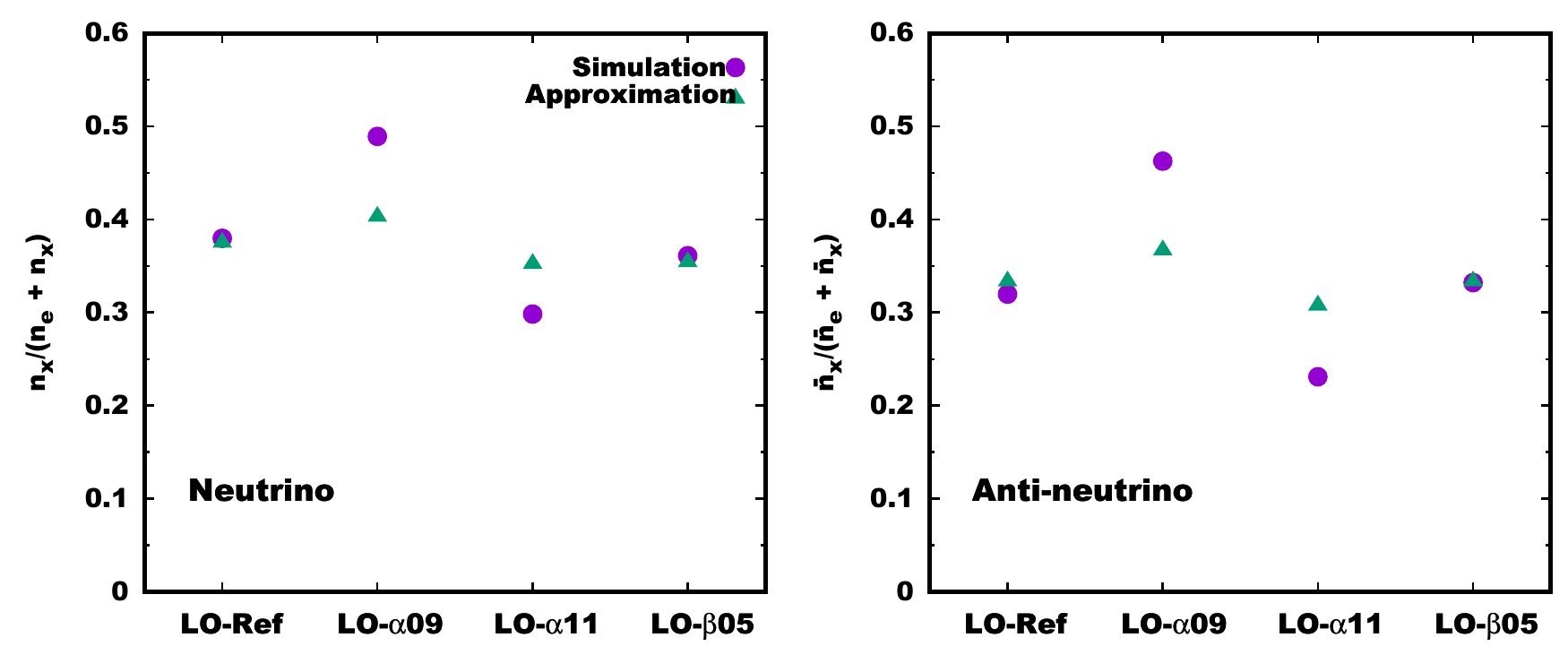}
   \caption{Comparison between FFC simulations (purple filled circles) and the approximate scheme (green triangles). The left panel displays $\nu_x/(\nu_e + \nu_x)$ after the system reaches in quasi-steady states. In FFC simulations, we employ results of local simulations, and compute the spatial average in the region of $50 {\rm km} + 1 {\rm m} < R < 50 {\rm km} + 5 {\rm m}$, where the system establishes the quasi-steady state, at the end of our simulations. Right panel displays the result for antineutrinos.
}
   \label{graph_phenocheck}
\end{figure*}

\begin{figure}
   \includegraphics[width=\linewidth]{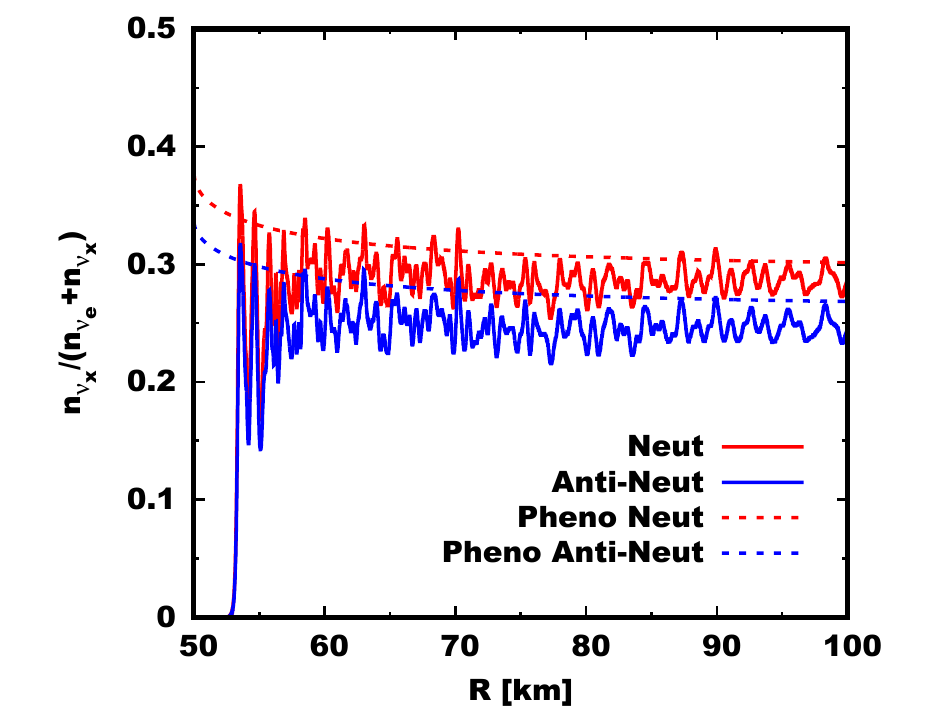}
   \caption{Radial profiles of $n_{\nu_{x}}/(n_{\nu_{e}} + n_{\nu_{x}})$ (red) and their antineutrinos (blue). The solid line shows their time-averaged profiles for GL-Ref model. The dashed line represents the results obtained by classical neutrino transport with our approximate scheme. See text for more details.
}
   \label{graph_compareRaddistriPheno}
\end{figure}

In the approximate method, we start with computing energy-integrated neutrino distributions but we leave the angular dependence in momentum space. For multi-angle neutrino transport sheme, this is straightforward. For approximate ones such as two-moment method, it is necessary to reconstruct full angular distributions of neutrinos from low angular moments. There are currently multiple options to do this. The most useful approach may be to use a so-called maximum-entropy completion \cite{1978JQSRT..20..541M,2021PhRvD.103l3012J,2022arXiv220608444R}, in which the full angular distributions are reconstructed so as to maximize the entropy under given zeroth and first angular moments. This approach can be adopted in both CCSN and BNSM simulations. For CCSN simulations, another approach proposed in our previous paper \cite{2021PhRvD.103l3025N} may offer a more accurate prescription, since the method is specifically desigined for CCSN. In this method, angular distributions of neutrinos, obtained by CCSN simulations with full Boltzmann neutrino transport, are fitted by two quadratic functions in a picewise manner, which allows us to reconstruct full angular distributions from the zeroth and first angular moments\footnote{The data table is available from \url{https://hirokinagakura.github.io/scripts/data.html}}.

Given energy-integrated neutrino distributions, we then make a rough estimation of the timescale of FFCs ($\tau_{\rm FFC}$) by using an empirical formula in \cite{2019ApJ...886..139N,2020PhRvR...2a2046M} with a minor extension,
\begin{equation}
\tau_{\rm FFC} = \frac{1}{c \sqrt{  - (\int_{G>0} G(\Omega) \frac{d \Omega}{4 \pi} ) (\int_{G<0} G(\Omega) \frac{d \Omega}{4 \pi} ) } },
\label{eq:FFCtime}
\end{equation}
where
\begin{equation}
G(\Omega) \equiv \sqrt{2} G_F \int_{0}^{\infty} \left( f_{{\rm ELN}} - f_{{\rm XLN}} \right) \frac{E_{\nu}^2 dE_{\nu}}{2 \pi^2}.
\end{equation}
In the expression, we write the speed of light explicitly; $\Omega$ represents the solid angle in momentum space, which specifies neutrino flight direction.

We then define an advection timescale ($\tau_{\rm adv}$) as
\begin{equation}
\tau_{\rm adv} \equiv \frac{R}{c}.
\label{eq:advTime}
\end{equation}
By using the two timescales, we define a new variable $q$ as,
\begin{equation}
q \equiv \min \left( \frac{ \tau_{\rm adv} }{ \tau_{\rm FFC} } , 1 \right),
\label{eq:advTime}
\end{equation}
which is a metric how large neutrino-mixings occur in the system. In the case with $q=1$, the timescale of FFC is shorter than the advection one, i.e., FFCs can be matured locally. For $q \ll 1$, the neutrinos would advect before FFC develops substantially. We control $p$ and $\bar{p}$ by using $q$; more specifically, the angular dependent $p$ and $\bar{p}$ is computed as,
\begin{equation}
p(\theta_{\nu}) = \bar{p}(\theta_{\nu}) = q \hspace{0.5mm} p_{\rm tmp}(\theta_{\nu}) + (1-q),
\label{eq:pFFCM}
\end{equation}
where
\begin{equation}
p_{\rm tmp} =
\begin{cases}
p_{\rm eq} & (\cos \theta_{\nu} \leq \cos \theta_{\nu (c)})\\
A \cos \theta_{\nu} + B & (\cos \theta_{\nu} > \cos \theta_{\nu (c)}),
\end{cases}
\label{eq:ptmpFFCM}
\end{equation}
with
\begin{equation}
  \begin{split}
& A = \frac{1-p_{\rm eq}}{1-\cos \theta_{\nu (c)}} \\
& B = \frac{p_{\rm eq} - \cos \theta_{\nu (c)}}{1-\cos \theta_{\nu (c)}}
  \end{split}
\label{eq:coef_ptmp}
\end{equation}
In the expression, $p_{\rm eq}$ denotes the survival probability for the case with flavor equipartition, i.e., $p_{\rm eq}=1/2$ and $1/3$ for two-flavor and three-flavor cases, respectively (see also Eqs.~\ref{eq:trapro_two}~and~\ref{eq:trapro_three}).

A few important remarks should be made here. First, the growth rate of FFC is only used to judge if FFCs can occur for given neutrino distributions. As long as the condition of $\tau_{\rm adv}>\tau_{\rm FFC}$ is satisfied, the growth rate has no influence on determining flavor conversions. Second, $q$ becomes zero in the case of no ELN-XLN crossing ($\tau_{\rm FFC} \rightarrow \infty$), which guarantees that no flavor conversions occur. Third, our method guarantees that ELN-XLN crossing disappears in the case of $q=1$, since ELN-XLN becomes zero in the angular region of $\cos \theta_{\nu} \leq \cos \theta_{\nu (c)}$. Fourth, the survival probability in our proposed method is a continuous functions of $\theta_{\nu}$ and $q$, which would be suited for sustaining stabilities of numerical simulations.

Another thing we do notice here is that our proposed method is capable of capturing an important property of FFCs found in our numerical simulations; the angular-averaged survival probability is different between neutrinos and antineutrinos. This is by virtue of the fact that we leave the angular dependence of $p$ and $\bar{p}$ in this method. As a result, both angular distributions of neutrinos and anti-neutrinos have direct influences on the total amount of flavor conversions, leading to different survival probabilities.

Although detailed inspections of the ability of the approximate scheme is postponed to another paper, we check the ability of the approximate scheme by comparing $n_{\nu_x}/(n_{\nu_e} + n_{\nu_x})$ for some selected models (LO-Ref, LO-$\alpha$09, LO-$\alpha$11, and LO-$\beta$05) to those obtained in numerical simulations (see Fig.~\ref{graph_phenocheck}). In numerical simulations, we compute the spatial average of $n_{\nu_x}/(n_{\nu_e} + n_{\nu_x})$ in the region of $50 {\rm km} + 1 {\rm m} < R < 50 {\rm km} + 5 {\rm m}$ at the end of each simulation\footnote{The spatially-averaged profile represent the quasi-steady state of FFCs, and they are almost identical to time-averaged ones in local simulations.}. We confirm that the results computed by the approximate method is in reasonable agreement with these local simulations.

We should mention two caveats, however. First, the approximate method has a relatively large deviation from numerical simulations for LO-$\alpha$09 and LO-$\alpha$11. This is due to a crude treatment of angular distribution of flavor conversion. For the sake of simplicity, we adopt a linear interpolation of $p_{\rm tmp}$, in the angular region of $\cos \theta_{\nu} > \cos \theta_{\nu (c)}$ (see Eq.~\ref{eq:ptmpFFCM}). However, numerical simulations suggest that FFCs in the angular region should be weaker (stronger) than those obtained by our approximate scheme for LO-$\alpha$09 (LO-$\alpha$11). It would be possible to improve the determination of $p_{\rm tmp}$ in the corresponding angular region, but work on improvements is currently underway. Second, our approximate scheme can not give an accurate estimation in cases that FFC is not the dominant mode of the instability. LO-$\bar{\beta}$0001 is such a case among our local simulations. This suggests that we need another parameter representing which mode (slow or fast) dominates the instability. We address these important issues in future work.

It is also important to test the ability of approximate scheme in global scale. One thing we need to notice here is that the asymptotic angular distribution of neutrinos are determined locally in the approximate scheme. However, the obtained distributions can not be the asymptotic state in global scale, since geometrical effects of neutrino advection have influence on them. We, hence, need to assess the approximate scheme in global scale. Here, we carry out a new simulation, in which we set the same initial condition of GL-Ref, and then run the simulation with turning off neutrino oscillations, meanwhile the approximate scheme is implemented. This simulation runs until the system reaches steady state ($1 {\rm ms}$). In Fig.~\ref{graph_compareRaddistriPheno}, we compare the results of time-averaged radial profile of $n_{\nu_{x}}/(n_{\nu_{e}} + n_{\nu_{x}})$ (for both neutrinos and antineutrinos) to those obtained from classical neutrino transport with approximate scheme. As shown in this figure, they are in reasonable agreement with each other (the error is less than $10 \%$) except for very inner region ($\lesssim 55 {\rm km}$). We note that the suppression of FFC at the inner region in GL-Ref is due to attenuation of the Hamiltonian (see Sec.~\ref{subsec:dynamifeat}), indicating that the deviation is not a matter of concern. This test illustrates the fidelity of the approximate scheme in global simulations.

\section{Summary}\label{sec:summary}
In this paper we present a systematic study of collective neutrino oscillation, paying special attention to fast neutrino-flavor conversion (FFC), by performing local ($\sim 10 {\rm m}$) and large-scale ($50 {\rm km}$) simulations in spherical symmetry. In large-scale simulations, we attenuate the neutrino Hamiltonian potential in a parametric manner so as to make the simulations tractable, and we extract physically meaningful features by combining these results with those in local simulations. Based on numerical results, we develop a novel approximate method to determine neutrino radiation field in quasi-steady state of FFC without solving QKE. The key findings and conclusions in the present study are summarized below.

\begin{enumerate}
\item Our proposed method (attenuating Hamiltonian potential in a parametric manner) has an ability to capture intsinc properties of collective neutrino oscillations in global scales, as consistent with our previous study \cite{2022arXiv220604097N} (NZv1). It is possible to get rid of spurious features due to the artificial prescription by a convergence study and comparisons to local simulations.
\item We find that the temporal variations of angular moments of neutrinos become mild, since incoherent variations are cancelled in the angular integration. This suggests that numerical results in low angular moment schemes as two-moment methods would underestimate the temporal variations of flavor conversion. On the other hand, the characteristic frequency is essentially the same among neutrinos in different angles and their angular moments (see Sec.~\ref{subsec:tempovari}).
\item Strong FFCs can occur even for the case with low growth rate of flavor conversion. This exhibits that the growth rate of flavor conversion is not a good metric to determine the total amount of flavor conversion in the non-linear phase. In fact, our results suggest that shallow ELN crossings can trigger large flavor conversions (see, e.g., GL-$\bar{\beta}$001$\xi$-2 and LO-$\bar{\beta}$001 models, whose results are presented in Sec.~\ref{subsec:modeldepe}).
\item When the ELN crossing is very shallow, the slow mode overwhelms the fast one in flavor conversion. This leads to a distinct property from FFCs; for instance, the angular structure of flavor conversion is remarkably different; see GL-$\bar{\beta}$0001$\xi$-1, GL-H-$\bar{\beta}$0001$\xi$-1, and LO-$\bar{\beta}$0001 models, whose results are discussed in Sec.~\ref{subsec:modeldepe}.
\item ELN-XLN angular distributions determine the overall characteristics of FFC dynamics. We confirm that FFCs are saturated when angular crossings disappear in the time-averaged ELN-XLN distributions. It should be stressed, however, that ELN-XLN distributions are not sufficient information to determine the total amount of flavor conversion.
\item The total amount of flavor conversion is less correlated with the growth rate but rather angular structures of neutrino distributions. It can be determined by how many neutrinos are contained in angular regions where flavor conversions occur vigorously (see the discussion at the end of Sec.~\ref{subsec:quasisteady}).
\item Our numerical simulations suggest that neutrinos propagating in the angle of $\cos \theta_{\nu} \lesssim \cos \theta_{\nu (c)}$ ($\theta_{\nu (c)}$ denotes the ELN-XLN angular crossing point) undergo strong flavor conversion in cases with forward-peaked angular distributions of neutrinos. On the other hand, FFCs in the region of $\cos \theta_{\nu} \gtrsim \cos \theta_{\nu (c)}$ tend to be less vigorous than those in $\cos \theta_{\nu} \lesssim \cos \theta_{\nu (c)}$. It should be noted, however, that FFCs in this region can be strong if the ELN crossing is $\cos \theta_{\nu (c)} \rightarrow 1$; see, e.g., the results of GL-$\alpha$09 and LO-$\alpha$09 models.
\item The guiding principle of our approximate method is to determine survival probability of neutrinos so as to eliminate ELN-XLN angular distributions. This can be realized by imposing a condition that neutrinos (and antineutrinos) in the region of $\cos \theta_{\nu} \lesssim \cos \theta_{\nu (c)}$ achieve the flavor equipartion, i.e., $p=\bar{p}=p_{\rm eq}$. 
\item We assess the ability of our approximate method by comapring the total amount of flavor conversion to those obtained from numerical simulations, that lends confidence to our method. We provide a recipe of the method in Sec.~\ref{subsec:FFCM} so as for other groups to implement the approximate method easily in their classical neutrino transport schemes.
\end{enumerate}

One of the important conclusions in the present study is that FFCs would radically change the neutrino-radiation field in CCSN and BNSM even if the ELN(-XLN) crossing obtained from classical neutrino transport is very shallow. It is an intriguing question how they affect fluid dynamics, neutrino signal, and nucleosynthesis. Addressing these issues requires accurate CCSN and BNSM modeling with incorporating feedback between neutrino transport, matter interactions, and flavor conversions (but see \cite{2022arXiv221002106F} for our recent study of impacts of flavor conversion on explosive nucleosynthesis in CCSN). This present study provides a feasible and resonable approach to tackle the issue by existing classical neutrino transport codes. Hopefully, the proposed method will serve to facilitate access to QKE and collective neutrino oscillations for the community of CCSN and BNSM theorists.

\section{Acknowledgments}
We are grateful to Sherwood Richers, Lucas Johns, Chinami Kato, George Fuller and Shoichi Yamada for useful comments and discussions. The numerical simulations are carried out by using "Fugaku" and the high-performance computing resources of "Flow" at Nagoya University ICTS through the HPCI System Research Project (Project ID: 210050, 210051, 210164, 220173, 220047), and XC50 of CfCA at the National Astronomical Observatory of Japan (NAOJ). For providing high performance computing resources, Computing Research Center, KEK, and JLDG on SINET of NII are acknowledged. This work is supported by High Energy Accelerator Research Organization (KEK). MZ is supported by a JSPS Grant-in-Aid for JSPS Fellows (No. 22J00440) from the Ministry of Education, Culture, Sports, Science, and Technology (MEXT) in Japan.
\bibliography{bibfile}

\end{document}